\documentclass[aps,prl,twocolumn,superscriptaddress,showpacs]{revtex4}
\usepackage{graphicx}
\usepackage{graphicx,epstopdf,color}
\usepackage{amsfonts}
\usepackage{amsmath,amssymb,mathrsfs}
\usepackage{bm}
\usepackage{float}
\usepackage{pst-grad}
\usepackage{dcolumn}
\pacs{74.70.Xa, 72.15.-v}

\begin{document}
\def \FeTeSe{Fe$_{1+\delta}$Te$_{1-x}$Se$_x$}
\def \FeTe{Fe$_{1+\delta}$Te}
\def \Ba122{BaFe$_2$As$_2$}
\def \Ca122{CaFe$_2$As$_2$}
\def \CoBa122{Ba(Fe$_{1-x}$Co$_x$)$_2$As$_2$}
\def \KBa122{Ba$_{1-x}$K$_x$Fe$_2$As$_2$}
\def \LSCO{La$_{2-x}$Sr$_x$CuO$_4$}
\def \Sr327{Sr$_3$Ru$_2$O$_7$}
\def \Tc{$T_c$}
\def \Ts{$T_{s}$}
\def \TN{$T_N$}
\def \AF{antiferromagnetic}

\title{Ubiquitous signatures of nematic quantum criticality in optimally doped Fe-based superconductors}

\author{Hsueh-Hui Kuo}
\affiliation{Stanford Institute for Materials and Energy Sciences, SLAC National Accelerator Laboratory,\\ 2575 Sand Hill Road, Menlo Park, CA 94025, USA}
\affiliation{Geballe Laboratory for Advanced Materials and Department of Materials Science and Engineering, Stanford University, USA}

\author{Jiun-Haw Chu}
\affiliation{Stanford Institute for Materials and Energy Sciences, SLAC National Accelerator Laboratory,\\ 2575 Sand Hill Road, Menlo Park, CA 94025, USA}
\affiliation{Geballe Laboratory for Advanced Materials and Department of Applied Physics, Stanford University, USA}

\author{Johanna C. Palmstrom}
\affiliation{Stanford Institute for Materials and Energy Sciences, SLAC National Accelerator Laboratory,\\ 2575 Sand Hill Road, Menlo Park, CA 94025, USA}
\affiliation{Geballe Laboratory for Advanced Materials and Department of Applied Physics, Stanford University, USA}

\author{Steven A. Kivelson}
\affiliation{Stanford Institute for Materials and Energy Sciences, SLAC National Accelerator Laboratory,\\ 2575 Sand Hill Road, Menlo Park, CA 94025, USA}
\affiliation{Geballe Laboratory for Advanced Materials and Department of Physics, Stanford University, USA}

\author{Ian R. Fisher}
\affiliation{Stanford Institute for Materials and Energy Sciences, SLAC National Accelerator Laboratory,\\ 2575 Sand Hill Road, Menlo Park, CA 94025, USA}
\affiliation{Geballe Laboratory for Advanced Materials and Department of Applied Physics, Stanford University, USA}

\begin{abstract}
A key actor in the conventional theory of superconductivity is the induced interaction between electrons mediated by the exchange of virtual collective fluctuations,  originally phonons. Other collective modes that can play the same role, especially spin-fluctuations, have been widely discussed in the context of high-temperature and heavy Fermion superconductors. The strength of such collective fluctuations is measured by the associated susceptibility. Here we use differential elastoresistance measurements on five optimally doped Fe-based superconductors to reveal that a diverging \emph{nematic} susceptibility appears to be a generic feature in the optimal doping regime of these materials. The observation motivates consideration of the effects of nematic fluctuations on the superconducting pairing interaction in this family of compounds, and possibly beyond.
\end{abstract}

\maketitle

A growing body of evidence suggests the possibility of an intimate connection between electronic nematic phases \cite{Fradkin_2010} and high-temperature superconductivity. However, it is currently unclear to what extent there is any causal relationship between nematic fluctuations and superconductivity. Strongly anisotropic electronic phases have been found in the underdoped regime of both cuprate \cite{Ando_2002, Aharon_2003, Hinkov_2008, Daou_2010, Steve_RMP} and Fe-based \cite{Chuang_2010, Chu_2010, Tanatar_2010, Dusza_2011, Yi_2011, Lu_2014} high-temperature superconductors. For underdoped Fe-based systems, recent measurements of the elastoresisistance \cite{JH_2012,HH_2013,HH_2014}, Raman spectroscopy\cite{Gallais_2013,Blumberg_2016,Hackl_2016}, and elastic moduli \cite{Yoshizawa_2012,Boehmer_2014} for the representative electron-doped system Ba(Fe$_{1-x}$Co$_x$)$_2$As$_2$ reveal a divergence of the electronic nematic susceptibility upon approach to the tetragonal-to-orthorhombic structural phase transition, definitively establishing that the phase transition is driven by electronic correlation.  For the cuprates, recent x-ray diffraction \cite{Chang_2012,Ghiringhelli_2012, Blackburn_2013, Tranquada_1995} and NMR \cite{Wu_2011} measurements have revealed evidence for short-range charge density wave order in ``underdoped'' crystals. Although details of the charge ordered state(s) are still being established, these initial observations have at least motivated discussion of a possible ``vestigial'' nematic order \cite{Nie_2014}. Perhaps significantly, in the phase diagrams of both families of compounds, optimal doping is located close to putative quantum critical points \cite{Kivelson_1998, Ramshaw_2014, Analytis_2014} which potentially have a nematic character \cite{JH_2012,Nie_2014,Fujita_2014a}. 

From a theoretical perspective, recent treatments indicate that nematic quantum criticality, (i.e. quantum critical fluctuations caused by proximity to a nematic quantum critical point) can provide an enhancement of the existing pairing interaction. In particular, a pure nematic phase does not break the translational symmetry of the original crystal lattice; consequently the $q$=0 nematic  fluctuations enhance $T_c$ in all symmetry channels \cite{Metlitski_2014, Lederer_2014, Maier_2014}. It is therefore of considerable interest to empirically establish whether nematic fluctuations are a characteristic feature of optimally doped high temperature superconductors, as well as to probe the extent to which these nematic fluctuations show intrinsically quantum behavior. In the present paper, we show that this is the case for Fe-pnictide and chalcogenide superconductors by considering the representative materials Ba(Fe$_{1-x}$Co$_x$)$_2$As$_2$ and Ba(Fe$_{1-x}$Ni$_x$)$_2$As$_2$ (i.e. electron-doped ``122''), Ba$_{1-x}$K$_x$Fe$_2$As$_2$ (hole-doped), BaFe$_2$(As$_{1-x}$P$_x$)$_2$ (isovalent substitution) and FeTe$_{1-x}$Se$_x$ (``11''). Furthermore, we find that the nematic susceptibility obeys a simple Curie-Weiss power law for all five optimally doped Fe-based superconductors over a wide temperature range. For the electron and hole-doped 122 pnictides a sub-Curie-Weiss deviation was observed at low temperatures, which we tentatively attribute to an enhanced sensitivity to disorder in a quantum critical regime.

Nematic order couples linearly to anisotropic strain of the same symmetry. Consequently, the nematic susceptibility of a material can be measured by considering the electronic anisotropy that is induced by anisotropic in-plane strain. In the regime of infinitesimal strains, all forms of electronic anisotropy are linearly proportional. Hence, the rate of change of resistivity anisotropy with respect to anisotropic strain, defined in the limit of vanishing strain, is linearly proportional to the nematic susceptibility\cite{constant}. The proportionality constant depends on microscopic physics, but away from any quantum critical point does not contain any singular behavior. Consequently, the induced resistivity anisotropy reveals the essential divergence of the nematic susceptibility upon approach to a thermally-driven nematic phase transition \cite{HH_2013, HH_2014}.

\begin{figure}
\begin{center}
\includegraphics[width=8.5cm]{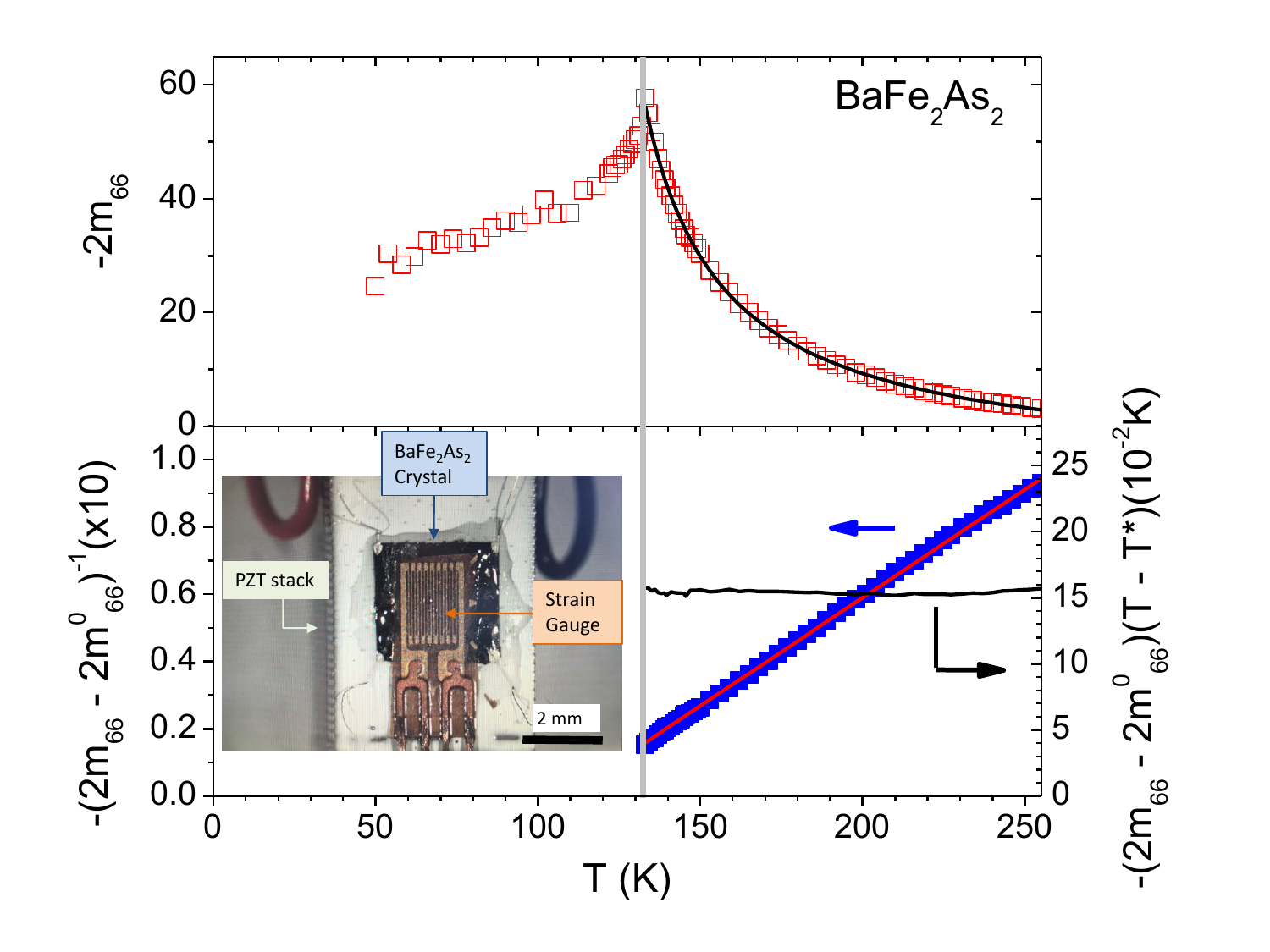} 
\caption{\textbf{Temperature dependence of the $B_{2g}$ elastoresistance of BaFe$_2$As$_2$.} The data follow a Curie-Weiss behavior, which is the anticipated mean-field temperature dependence of the nematic susceptibility of a material approaching a thermally driven nematic phase transition \cite{JH_2012,HH_2013,HH_2014}. Upper panel shows $-2m_{66}$, proportional to the nematic susceptibility $\chi_{N(B_{2g})}$. Black line shows the Curie-Weiss fit. The quality of fit can be better appreciated by considering the inverse susceptibility, $-(2m_{66}-2m_{66}^0)^{-1}$ which is perfectly linear (left axis of lower panel; fit shown by red line), and the Curie constant $-(2m_{66}-2m_{66}^0)^{-1}(T-T^*)$ (right axis of lower panel), which is independent of temperature. The Weiss temperature obtained from the Curie-Weiss fit, which gives the bare mean field nematic critical temperature, yields a value $T^*$ = 109 $\pm$1.4 K. Coupling to the lattice renormalizes the critical temperature, leading to a  nematic-structural phase transition at $T_s$ = 134 K (vertical gray line). Inset shows a photograph of a square crystal glued on a PZT piezoelectric stack for differential elastoresistance measurements using Montgomery's geometry. Four electrical contacts were made at the corners and a strain gauge was glued on the top surface\cite{SOM}. } 
\end{center}
\end{figure}
\begin{figure*}
\begin{center}
\includegraphics[width=17cm]{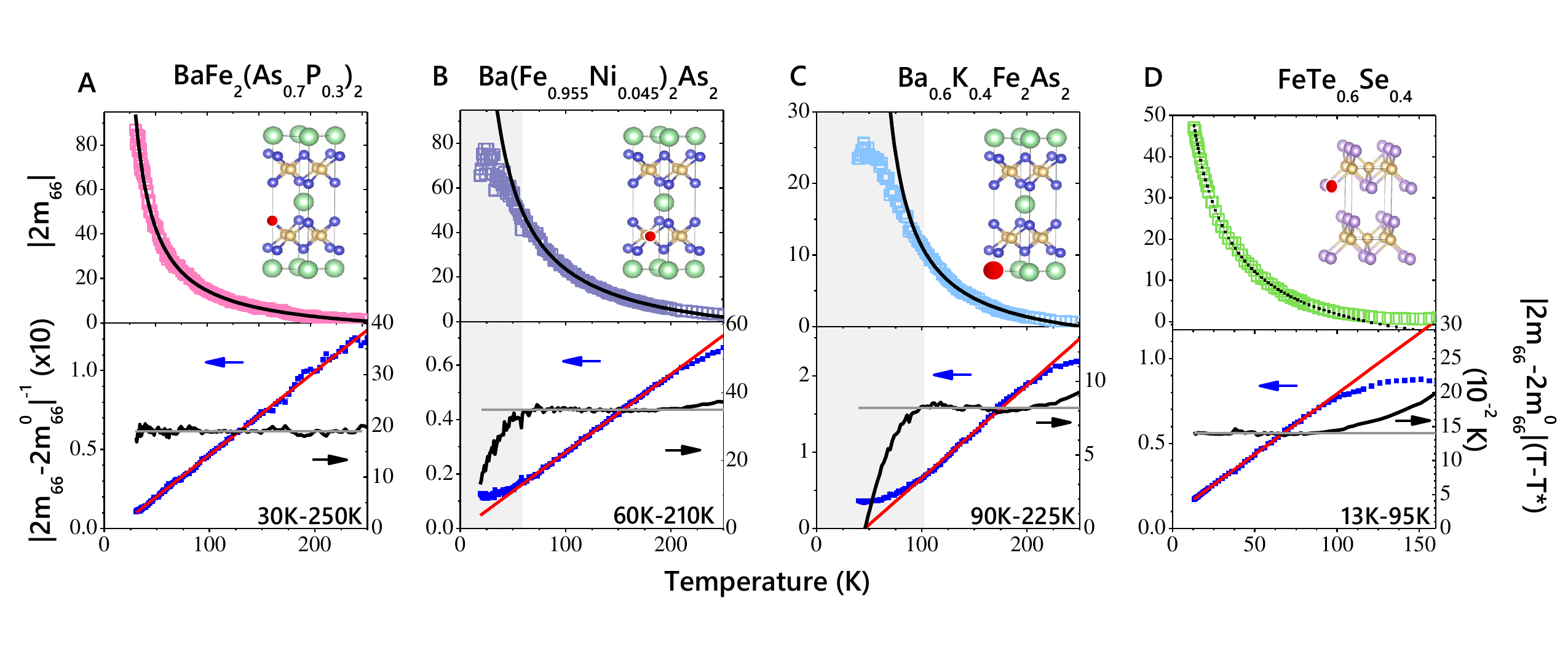}[h] 
\caption{\textbf{Divergence of the $B_{2g}$ elastoresistance $2m_{66}$ of several different families of optimally doped Fe pnictide and chalcogenide superconductors.} (A) optimally doped BaFe$_2$(As$_{0.7}$P$_{0.3}$)$_2$ (isovalent substitution),  (B) optimally doped Ba(Fe$_{0.955}$Ni$_{0.045}$)$_2$As$_2$ (electron doped), (C) optimally doped Ba$_{0.58}$K$_{0.42}$Fe$_2$As$_2$ (hole doped), and (D) optimally doped FeTe$_{0.58}$Se$_{0.42}$. Insets indicate the dopant site (red) in the respective unit cells of each material. Comparable magnitude induced resistivity anisotropies are observed for each case\cite{SOM}. Upper panels show $|2m_{66}|$, whereas lower panels show $|(2m_{66}-2m_{66}^0)|^{-1}$ (left axes of lower panels, blue symbols) and $|(2m_{66}-2m_{66}^0)|(T-T^*)$ (right axes of lower panels, black curves). Black(upper panels) and red(low panels) lines shows fits to Curie-Weiss behavior of $m_{66}$ and $|(2m_{66}-2m_{66}^0)|^{-1}$ respectively. Grey horizontal lines (low panels) shows the average values of  $|(2m_{66}-2m_{66}^0)|(T-T^*)$ in the fitting temperature range. Regions of deviation from Curie-Weiss behavior in (B) and (C) are indicated by gray shaded regions. For (A) and (B), $2m_{66}$ is negative. For (C) and (D), $2m_{66}$ is positive. Fit parameters are listed in \cite{SOM,Dimension}.} 
\end{center}
\end{figure*} 
For a tetragonal material, nematic order has either $d_{xy}$ symmetry (i.e. the $B_{2g}$ irreducible representation of the $D_{4h}$ point group, corresponding to nematic order oriented along a nearest-neighbor Fe-Fe bond -- the [110] or [1{\=1}0] crystal axes), or $d_{x^2-y^2}$ symmetry ($B_{1g}$, corresponding to the [100] or [010] crystal axes). Anisotropic strains with appropriate symmetry are then $\epsilon_{6} = (\epsilon_{xy}+\epsilon_{yx})/2$ and $\epsilon_{1}-\epsilon_{2} = \epsilon_{xx}-\epsilon_{yy}$ respectively. The strain-induced changes in resistivity can be described using the dimensionless elastoresistivity tensor, $m_{ij}$ \cite{HH_2013}: 

\begin{equation}
(\Delta\rho/\rho)_i = \displaystyle\sum_{j=1}^{6} m_{ij}\epsilon_j
\end{equation}
where $1=xx$, $2=yy$ etc. Consequently, the corresponding components of the nematic susceptibility tensor are given by: 

\begin{eqnarray}
& \chi_{N(B_{2g})} = c \times 2m_{66}\\ 
& \chi_{N(B_{1g})} = c' \times (m_{11}-m_{12})
\end{eqnarray}
where $c$ and $c'$ are proportionality constants, which depend on microscopic physics\cite{constant}. 

We measure elastoresistivity by applying an in-situ tunable anisotropic strain using a piezoelectric PZT stack. A square plate sample (typical dimension 750x750x20 $\mu$m) is glued on the side wall of the PZT stack using a commercial two part epoxy. The PZT stack deforms when a voltage is applied and hence strains the sample glued on top of it\cite{Strain}. The amount of strain can be measured by a strain gauge glued either on the backside of the PZT stack or on the top surface of a larger sample(the latter case enabling a full determination of the strain transmission\cite{SOM}). The in-plane resistivity tensor of the sample is measured via the Montgomery technique, with electrical contacts made at the four corners of the square sample \cite{SOM}. Representative data taken using this new technique are shown in Figure 1 for the specific case of BaFe$_2$As$_2$. As has been  previously demonstrated \cite{HH_2013, HH_2014}, the data can be fit very well by a Curie-Weiss temperature dependence:

\begin{equation}
2m_{66} = 2m_{66}^0 + \lambda/[a(T-T^*)].
\end{equation}

In Figure 2 we show $B_{2g}$ elastoresistance data for a range of optimally doped materials, including BaFe$_2$(As$_{1-x}$P$_{x}$)$_2$ (isovalently substituted), Ba(Fe$_{1-x}$Ni$_x$)$_2$As$_2$ (electron-doped), Ba$_{1-x}$K$_x$Fe$_2$As$_2$ (hole-doped) and FeTe$_{1-x}$Se$_x$. In all cases, $2m_{66}$ rises strongly with decreasing temperature, with comparably large values for each compound. The observation of such an effect for this wide variety of ways of doping, including the case of the iron chalcogenide FeTe$_{0.6}$Se$_{0.4}$, is highly suggestive that a divergence of the nematic susceptibility in the $B_{2g}$ channel is a generic feature of optimally doped Fe-based superconductors. This is our main result. Regardless of microscopic models, from a purely empirical perspective it is apparent that the optimally-doped superconductor is born out of an electronic state that is characterized by strongly fluctuating orientational (nematic) order in this specific symmetry channel. This result is especially notable for FeTe$_{0.6}$Se$_{0.4}$ given that the orientation of both the magnetic ordering wave vector and the in-plane component of the structural distortion of undoped FeTe is 45 degrees away from the orientation of those in the iron arsenide materials.

\begin{figure*}
\begin{center}
\includegraphics[width=17cm]{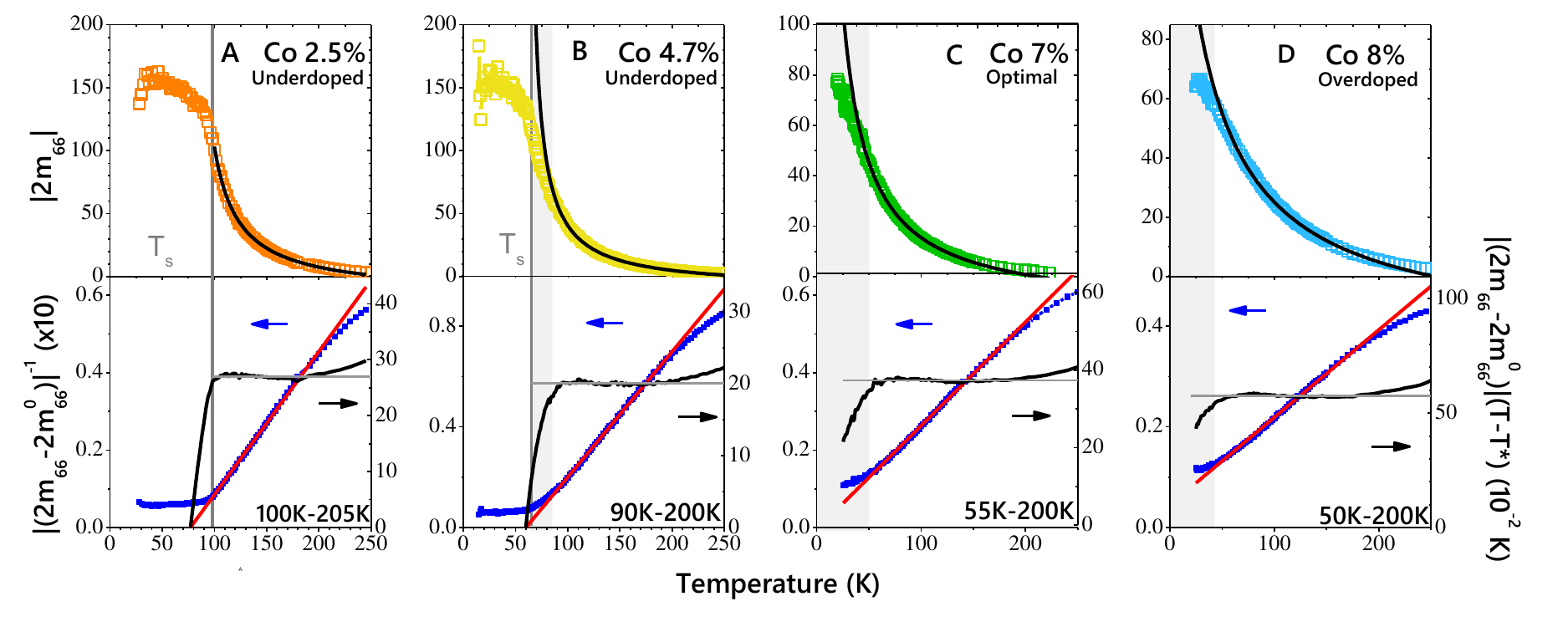} 
\caption{\textbf{Variation of the $B_{2g}$ elastoresistance of Ba(Fe$_{1-x}$Co$_{x}$)$_2$As$_2$ for four representative compositions.} (A,B) underdoped compositions Ba(Fe$_{0.975}$Co$_{0.025}$)$_2$As$_2$ and Ba(Fe$_{0.953}$Co$_{0.047}$)$_2$As$_2$; (C) optimal-doped Ba(Fe$_{0.93}$Co$_{0.07}$)$_2$As$_2$; and (D) overdoped Ba(Fe$_{0.92}$Co$_{0.08}$)$_2$As$_2$. The heavily underdoped composition is very well described by a Curie-Weiss temperature dependence over the entire temperature range(black lines in upper panels). For the compositions near optimally-doping, $2m_{66}$ can be well fit by a Curie-Weiss T-dependence at high temperatures. Below a characteristic temperature scale(different for each compositions, indicated by shade gray region), a strong downward deviation from Curie Weiss behavior is observed, also seen in the inverse susceptibility  (upward curvature) and in $|(2m_{66}-2m_{66}^0)|(T-T^*) \propto \chi_N(T-T^*)$ (strong downturn), which are shown in the lower panels. The deviation from Curie-Weiss behavior is the strongest at the optimal doping, and diminishes on either side of the phase diagram.  Curie-Weiss fit parameters for each composition are listed in \cite{SOM}.} 
\end{center}
\end{figure*}

The $m_{66}$ coefficient of optimally doped BaFe$_2$(As$_{0.68}$P$_{0.32}$)$_2$ (Fig. 2A) follows a perfect Curie-Weiss temperature dependence from $T = 250K$ (the highest temperature where strain can be effectively transmitted by the epoxy) down to $T_c$ (below which temperature the resistance is zero), with a Weiss temperature ($T^*$) close to zero. For FeTe$_{0.6}$Se$_{0.4}$, the $m_{66}$  coefficient can be perfectly fitted with Cure-Weiss down to $T_c$  but shows a significant deviation for temperatures greater than 100K, possibly related to the loss of quasi particle coherence due to its extremely small Fermi energy as observed in photoemission spectroscopy and transport\cite{Baumberger_2010,Behnia_2011}. Intriguingly, for the electron and hole doped ‘122’ pnictides, a downward deviation from Curie-Weiss behavior is observed below a characteristic temperature that is different for each of the materials. The quality of fit to the Curie-Weiss functional form for BaFe$_2$(As$_{0.68}$P$_{0.32}$)$_2$ in comparison to all of the other optimally doped compositions studied can be readily appreciated if the data are plotted on log axes (Fig. S11). Only the data for the P-substituted system can be fit by a single power law [$|(2m_{66}-2m_{66}^0)| \sim (T-T^*)^{- \gamma}$] over the entire temperature range, yielding $\gamma=0.985 \pm 0.005$ and $T^*=11.7 \pm 3.1$. 

A slightly overdoped BaFe$_2$(As$_{0.64}$P$_{0.36}$)$_2$ sample has also been measured; the elastoresistivity coefficient $m_{66}$ of this sample can also be well fitted by Curie-Weiss temperature dependence over the entire temperature range, with a negative Weiss temperature $T^*=-11.5 \pm 2.3$(Fig. S14). The small value of the Weiss temperature observed in BaFe$_2$(As$_{1-x}$P$_{x}$)$_2$ near optimally doping, and the fact that it crosses zero as doping increases, motivates consideration of the nematic susceptibility in the context of a quantum critical point. An Ising nematic phase transition in an insulator would generally be expected to have dynamical exponent $z=1$, so the effective dimensionality of the system would be $d+z$ = 3+1. As the materials in question are metallic, the situation is more complicated \cite{Sri_2013}; the Hertz-Millis paradigm is expected to break down in the case of electronic nematic order in metallic systems, for which Landau damping associated with the gapless Fermions leads to a larger value of the dynamical exponent, $z$ = 3. Although the analysis of this problem is not straightforward, both analytic \cite{Hartnoll_2014} and numerical studies \cite{Lederer_2015} indicate that the nematic susceptibility at criticality should diverge in proportion to 1/T (up to possible logarithmic corrections), consistent with the observed behavior of the $B_{2g}$ elastoresistance of optimally doped BaFe$_2$(As$_{1-x}$P$_{x}$)$_2$.

To gain insight into the physical origin of the low temperature downward deviation from Curie-Weiss behavior in the electron and hole doped pnictides, we consider the evolution of the $B_{2g}$ elastoresistance as a function of composition for the specific case of Co-doped BaFe$_2$As$_2$. Figure 3 A-D show the progression of $2m_{66}$ as a function of Co doping for several representative compositions (more were measured, see \cite{SOM}), from underdoped (A,B), through optimal doping (C), to the overdoped regime (D).  For the heavily underdoped compositions(x = 0.025), $2m_{66}$ can be well described by mean field Curie-Weiss T-dependence down to $T_s$. It has been previously established that this behavior is essentially independent of disorder, at least comparing undoped BaFe$_2$As$_2$ with different Residual Resistance Ratios (RRR), and Co and Ni substituted crystals with the same $T_N$ \cite{HH_2014}. The Weiss temperature $T^*$ extracted from high-temperature fits to Curie-Weiss behavior crosses zero close to optimal doping (Fig. 4). However, for lightly underdoped Ba(Fe$_{0.553}$Co$_{0.047}$)$_2$As$_2$, a downward  deviation from mean-field behavior at low temperatures begins to be noticeable, and becomes more pronounced for optimally doped Ba(Fe$_{0.93}$Co$_{0.07}$)$_2$As$_2$; similar deviations are observed in optimally Ni and K doped BaFe$_{2}$As$_2$. A similar effect has been observed in recent measurements of the sheer modulus via three-point bending experiments  \cite{Boehmer_2014}. The deviation from Curie-Weiss behavior diminishes again as the doping is further increased:  for the overdoped composition (Fig. 3D) the data can be fit to Curie-Weiss to a lower temperature than for optimal doping, and the magnitude of the deviation below $\sim$ 45 K is smaller than for optimal doping. $2m_{66}$ of four additional Co dopings has also been measured, showing a similar non-monotonic doping dependence of the deviation from Curie-Weiss behavior. The fact that the effect is maximal near optimal doping where $T^* \sim 0$, suggests that it is associated with proximity to the putative nematic quantum phase transition.

P-substituted BaFe$_2$As$_2$ is the ``cleanest'' of all of the known 122 Fe-pnictide families, evidenced by the fact that quantum oscillations can be observed across the phase diagram \cite{Jim_QO, Shishido_2010}. The deviation from Curie-Weiss behavior for optimally doped compositions of all other dopants in BaFe$_2$As$_2$ suggests that disorder plays an important role in the quantum critical regime. All forms of quenched disorder produce locally anisotropic effective strains, which thus couple to the orientation of the nematic order; this is ``random-field'' disorder\cite{Nie_2014}. Analysis of the random-field Ising model yields several generically expected effects of random field disorder, including suppression of the nematic susceptibility below mean field expectations for a clean system, and, for the case of a quantum phase transition, the enhanced sensitivity of quantum critical phenomena to disorder.\cite{SOM}. 

The temperature dependence of the nematic susceptibility can also be extracted from measurements of the elastic moduli \cite{Boehmer_2014}. The two measurements (elastoresistance and elastic moduli) are in broad agreement, for example in terms of the Curie-Weiss T-dependence of $\chi_N$ for underdoped compositions of Ba(Fe$_{1-x}$Co$_x$)$_2$As$_2$, and also in terms of the deviation from Curie-Weiss behavior near optimal doping. However, there is an important distinction in terms of the relative magnitude of the measured quantities as a function of doping. In particular, the normalized lattice softening $[c_{66}^0-c_{66}]/c_{66}^0$ extracted in \cite{Boehmer_2014} monotonically decreases in magnitude and extent in temperature as a function of $x$. In contrast, the quantity $|(2m_{66}-2m_{66}^0)| = c \chi_N$ initially increases with $x$, peaking for lightly underdoped compositions $x\simeq 0.05$.	 The apparent enhancement of the elastoresistance 2$m_{66}$ over the softening of the elastic modulus for compositions near optimal doping is potentially related to renormalization of the quasiparticle effective mass in the quantum critical regime, as has been seen in P-substituted BaFe$_2$As$_2$ \cite{Analytis_2014}. Such a mass renormalization is an expected consequence of nematic quantum critical fluctuations\cite{Sri_2013}. Significantly, it is precisely these low-energy quasiparticles that are also involved in the eventual superconductivity.

\begin{figure}
\includegraphics[width=8.5cm]{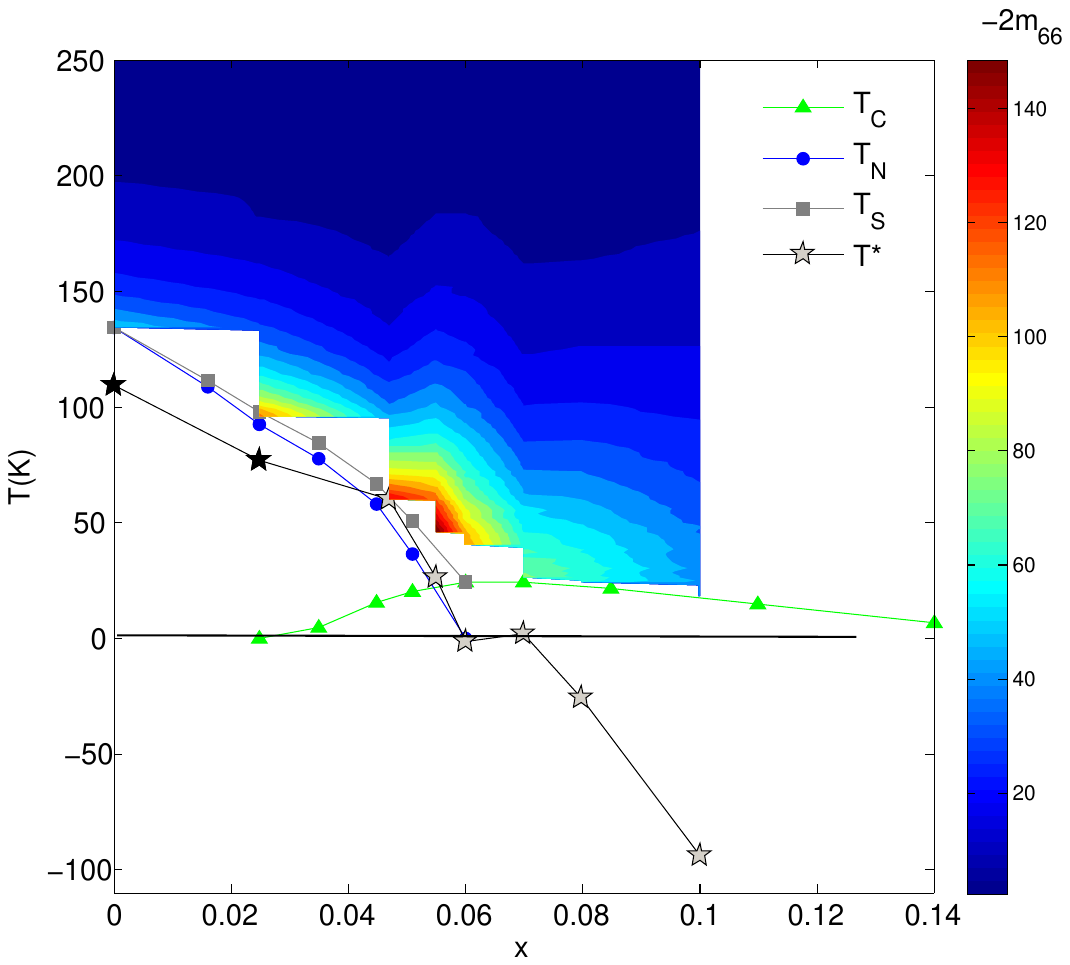} 
\caption{\textbf{Phase diagram of Ba(Fe$_{1-x}$Co$_x$)$_2$As$_2$, showing the variation of $2m_{66}$ in the $x-T$ plane (color scale).} Squares, circles and triangles indicate T$_s$, T$_N$, and T$_c$ respectively. Stars indicate the bare mean-field nematic critical temperatures, $T^*$, extracted from the Curie-Weiss fits of $2m_{66}$ above the temperature at which disorder effects suppress the nematic susceptibility (see text; light grey symbols are used for the cases for which deviations from Curie Weiss behavior are observed at low temperatures). As has been previously determined via longitudinal elastoresistance measurements\cite{JH_2012}, but established here by the full $B_{2g}$ differential elastoresistance, the Weiss temperature, $T^*$ goes through zero as a function of $x$ close to optimal doping. Color scale shows the magnitude of $-2m_{66}$, which peaks between lightly underdoped to optimal doped compositions.} 
\end{figure}

The temperature dependence of the nematic susceptibility can also be extracted from measurements of the elastic moduli \cite{Boehmer_2014}. The two measurements (elastoresistance and elastic moduli) reveal broad agreement, for example in terms of the Curie-Weiss T-dependence of $\chi_N$ for underdoped compositions of Ba(Fe$_{1-x}$Co$_x$)$_2$As$_2$, and also in terms of the deviation from Curie-Weiss behavior near optimal doping. However, there is an important distinction in terms of the relative magnitude of the measured quantities as a function of doping. In particular, the normalized lattice softening $[c_{66}^0-c_{66}]/c_{66}^0$ extracted by Boehmer \emph{et~al.} \cite{Boehmer_2014} monotonically decreases in magnitude and extent in temperature as a function of $x$. However, as can be seen from inspection of Figure 3, the quantity $|(2m_{66}-2m_{66}^0)| = c \chi_N$ initially increases with $x$, peaking for optimally doped compositions. The apparent enhancement of the elastoresistance 2$m_{66}$ over the softening of the elastic modulus for compositions near optimal doping is potentially related to renormalization of the quasiparticle effective mass in the quantum critical regime, as has been seen for instance in P-substituted BaFe$_2$As$_2$ \cite{Analytis_2014}. Such a mass renormalization is an expected consequence of nematic quantum critical fluctuations\cite{Sri_2013}. Significantly, it is precisely the low-energy quasiparticles that determine the elastoresistance that are also involved in the eventual superconductivity. 

\section{Acknowledgments}
The authors thank S. Raghu, S. Lederer, E. M. Spanton for helpful discussions. Johanna Palmstrom is supported by a Gabilan Stanford Graduate Fellowship and a National Science Foundation Graduate Research Fellowship. This work was supported by the DOE, Office of Basic Energy Sciences, under Contract No. DE-AC02-76SF00515 and by the National Science Foundation Graduate Research Fellowship under Grant No. DGE-114747.

\section*{Materials and Methods}

\subsection*{Elastoresistivity coefficients}

Elastoresistivity relates changes of resistivity, $\rho$, of a solid conductor to strains that it experiences. We define the change in resistivity with respect to a strain $\epsilon$ as \cite{HH_2013, Max_Sym}:
\begin{eqnarray}
\Delta\rho_{\alpha\beta} = \rho_{\alpha\beta}(\epsilon)-\rho_{\alpha\beta}(\epsilon = 0)
\end{eqnarray}
And the normalized resistivity change as \cite{HH_2013, Max_Sym}:
\begin{eqnarray}
(\Delta\rho/\rho)_{\alpha\beta} = \frac{\Delta\rho_{\alpha\beta}}{\sqrt{\rho_{\alpha\alpha}}\sqrt{\rho_{\beta\beta}}}
\end{eqnarray} 
Since resistivity and strain are both 2nd rank tensors, the elastoresistivity, $M$, is a 4th rank tensor defined by the following expression \cite{Max_Sym}:
\begin{eqnarray}
(\Delta\rho/\rho)_{\alpha\beta} = \sum_{\gamma\delta}M_{\alpha\beta\gamma\delta}\epsilon_{\gamma\delta}\\
\alpha ,\beta ,\gamma ,\delta = x,y,z
\end{eqnarray}
In the absence of a magnetic field, the Onsager reciprocal relation holds for the electrical conductivity, and the associated symmetry properties of the resistivity and strain tensors allow us to more conveniently express them as six component vectors, such that
\begin{eqnarray}
(\Delta\rho/\rho)_{k} = \sum^6_{l=1}m_{kl}\epsilon_l
\end{eqnarray}
where $1 = xx, 2 = yy, 3 = zz, 4 = yz, 5 = zx, 6 = xy$. Using this Voigt notation, the elastoresistivity is represented by a pseudo-second rank tensor. For a tetragonal crystal, the point group symmetry allows us to further simplify the elasotresistivity tensor:
\begin{eqnarray}
m_{ik} = \left( \begin{array}{cccccc}
m_{11} & m_{12} & m_{13} & 0 & 0 & 0 \\
m_{12} & m_{11} & m_{13} & 0 & 0 & 0 \\
m_{31} & m_{31} & m_{33} & 0 & 0 & 0 \\
0 & 0 & 0 & m_{44} & 0 & 0\\
0 & 0 & 0 & 0 & m_{44} & 0\\
0 & 0 & 0 & 0 & 0 & m_{66}\\
\end{array} \right)
\end{eqnarray}
Here the six compenent vectors describing the resistivity change and the strain are defined with respect to the $a$, $b$, and $c$ crystollographic axes which are aligned with the $x$, $y$, and $z$ Cartesian axes. The elastoresistivity coefficients will be written with respect to the crystollographic axes throughout this text. In the following sections, we describe how the elastoresistivity coefficients $(m_{11}-m_{12})$ and $m_{66}$ are related to the nematic susceptibility for the $B_{1g}$ and $B_{2g}$ symmetry channels, and how these coefficients can be measured. These ideas were first described in ref \cite{HH_2013}. 

\subsection*{Elastoresistivity coefficient and nematic susceptibility}

For a C$_4$ to C$_2$ phase transition the nematic order parameter is a Ising variable, and is proportional to the anisotropy of any equilibrium electronic properties\cite{Fradkin_2010}. The dimensionless resistivity anisotropy
\begin{eqnarray}
N = \frac{\rho_{xx} - \rho_{yy}}{(\rho_{xx} + \rho_{yy})}
\end{eqnarray}
does not purely measure the equilibrium electronic properties, but is determined by the anisotropy in the electronic structure \emph{and} the anisotropy in the scattering rate; the latter being a dynamical property of the system. Nevertheless, one can still express the nematic order parameter $\Psi$ as a function of the resistivity anisotropy $N$ and take a Taylor expansion for small value of $N$, for which the leading term is linear. Therefore, for small value of $N$, which applies to our measurements (which are all in the linear regime), it can be related to the nematic order parameter via a proportionality constant $N$ = $c\Psi$.
Strain with the same symmetry (i.e. $\epsilon_{xx} - \epsilon_{yy}$) couples linearly to $\Psi$, and hence can be considered as a symmetry breaking field conjugate to the electron nematic order parameter. Therefore we can define the nematic susceptibility $\chi_N$ as the quantity that relates the induced order parameter and symmetry breaking strain in the region of linear response. For a tetragonal crystal there are two possible symmetry channels in which nematic order can develop via a continuous phase transition. The two symmetry channels correspond to B$_{1g}$ and B$_{2g}$ irreducible representation of a tetragonal point group, which couples linearly to orthorhombic distortions oriented along [100]$_T$ and [110]$_T$ direction in the tetragonal lattice. Hence for a crystal oriented with [100]$_T$ along the $x$ direction (define by the PZT stack orientation):
\begin{eqnarray}
\chi_{N(B1g)} = \frac{d\Psi_{B1g}}{d(\epsilon_{xx} - \epsilon_{yy})}\vert_{\epsilon = 0} = c(m_{11}-m_{12}) 
\end{eqnarray}
Similarly, for a crystal oriented with [110]$_T$ along the $x$ direction:
\begin{eqnarray}
\chi_{N(B2g)} = \frac{d\Psi_{B2g}}{d\epsilon_{xy}}\vert_{\epsilon = 0} = c'm_{66}
\end{eqnarray}

Therefore the elastoresistivity coefficient is related to the bare nematic susceptibility via the same proportionality constant $c$. The proportionality constant $c$ depends on the microscopic details of the electronic structure, the precise definition given to the nematic order parameter, and, in some cases, the nature of the disorder scattering, and in general could be temperature and materials dependent. However, two experimental observations of iron based superconductors suggest that $c$ is a constant within a wide range of temperatures and independent to a certain degree of disorder\cite{HH_2014}. First, the temperature dependence of the elastoresistivity coefficient $m_{66}$ of the parent compounds \Ba122\ can be perfectly fitted by a Curie-Weiss function (Fig. 1 of main paper), which is exactly the same functional form of $\chi_N$ derived by a Ginzburg-Landau theory. Second, the elastoresistivity coefficient $m_{66}$ of \Ba122\ has been measured for crystals with different residual resistance ratios (RRR), which is a measure of crystalline disorder \cite{HH_2014}. It was found that $m_{66}$ is independent of the RRR of the sample. A similar result is found for underdoped compositions of Co- and Ni-doped BaFe$_2$As$_2$ which have the same Neel temperature\cite{HH_2014}. The perfect fitting to Curie-Weiss in these conditions strongly suggest that $c$ remains a constant over a wide range of temperatures and different degrees of disorder.

\subsection*{Elastoresistance Measurements using a modified Montgomery method}
The elastoresistivity coefficients $(m_{11}-m_{12})$ and $m_{66}$ can be determined from measurements of $\rho_{xx}$ and $\rho_{yy}$ as a function of anisotropic strain $\epsilon_{xx} - \epsilon_{yy}$\cite{HH_2013, Max_Sym}. Previously, we have demonstrated that anisotropic strain could be achieved by gluing thin crystals to the side surface of a PZT piezoelectric stack \cite{PiezoSt}. The resistivity  $\rho_{xx}$ and $\rho_{yy}$ were measured separately by gluing two long bar samples along the poling and the transverse directions of PZT stack. In this paper we introduce an improved technique by measuring $\rho_{xx}$ and $\rho_{yy}$ on a single square sample via a modified Montgomery method. The advantage of using a single square sample is twofold: by using a single sample we remove the uncertainties due to unequal strain experienced by two crystals; the square shape also eliminates possible errors due to geometric effects, ensuring that the strain transmission is equal along the $x$ and $y$ directions. In the remainder of this section, we describe the details of this new technique.

Single crystals were grown via standard self flux techniques, as described elsewhere \cite{JH_2009}. Square samples were cut from larger crystals such that the tetragonal c-axis was perpendicular to the plane of the thin squares, and the a-axis aligned either along the sides of the square (suitable for probing the B$_{1g}$ channel) or at 45 degrees (for B$_{2g}$). Anisotropic strain was achieved by gluing these thin crystals to the side surface of a PZT piezoelectric stack\cite{PiezoSt}, with the sides of the squares aligned along the principle $x$ and $y$ axes of the stack. By applying a positive voltage bias through the voltage leads, the PZT stack causes the sample to experience an anisotropic strain by expanding along the poling direction and contracting along the transverse direction. We define the poling direction of the PZT stack as the $y$-direction and the transverse direction as $x$-direction. The resistances along the $x$ and $y$-directions of a single sample were measured simultaneously while the voltage applied to the PZT stack was varied (typically for three full hysteresis loops between -50 V and +150 V) (Fig. S1). A photograph and illustration of the configuration of a typical experiment probing the elastoresistance of such a sample is shown in Figures S1 and S2. Current is sourced at a given frequency across the top two contacts A and B of the sample, while the resistance along the $x$-direction ($R_x$) is measured by the voltage drop across the bottom two contacts C and D. Simultaneously a second current at a separate, distinct frequency is sourced through the leftmost two contacts, C and D, while the resistance along the $y$-direction ($R_y$) is measured across the rightmost contacts, D and B. 

The resistivity along the $x$ and $y$-directions $\rho_{xx},\rho_{yy}$ can be calculated from the measured resistances $R_x,R_y$ using a modified Montgomery method if the sample side lengths ($L_{x}$ and $L_{y}$) and thickness ($L_{z}$) are known (Fig. S3)\cite{Montgomery, Shigue}. The square anisotropic crystal can be mapped to an isotropic equivalent sample with side lengths $L_{x}'$ and $L_{y}'$ using the expression:
\begin{eqnarray}
\frac{L_{y}'}{L_{x}'} \approx \frac{1}{2} \bigg[  \frac{1}{\pi}ln\frac{R_{y}}{R_{x}} + \sqrt{(\frac{1}{\pi}ln\frac{R_{y}}{R_{x}})^2+4}  \bigg]
\end{eqnarray}
where $R_{x}$ and $R_{y}$ are the measured resistances along the $x$ and $y$-axes respectively \cite{Montgomery, Shigue}. In the thin sample limit ($\frac{L_{z}}{(L_{x}L_{y})^\frac{1}{2}} <0.5$) this ratio can then be used to calculate the resistivities along the PZT axes according to the relations \cite{Montgomery, Shigue}:

\begin{eqnarray}
\rho_{xx} = \frac{\pi}{8}(\frac{L_{z}L_{y}}{L_{x}})(\frac{L_{x}'}{L_{y}'})R_{x}sinh(\frac{\pi L_{y}'}{L_{x}'})\\
\rho_{yy} = \frac{\pi}{8}(\frac{L_{z}L_{x}}{L_{y}})(\frac{L_{y}'}{L_{x}'})R_{y}sinh(\frac{\pi L_{x}'}{L_{y}'})
\end{eqnarray}

Neglecting geometric effects (ie small changes in $L_{x}$ and $L_{y}$ of the strained sample, which are discussed in the next section), $\rho_{xx}$ and $\rho_{yy}$ can be obtained for each applied strain at each temperature. Now we can express the experimentally measured resistivity change for a thin square sample experiencing $B_{1g}$ strain (e.g. it is glued so the square edges and the tetragonal a-axis of the crystal are aligned to the axis of the PZT stack) in terms of the elastoresistivity tensor by using the following expression:
\begin{eqnarray}
(\Delta\rho/\rho)_{xx} =\epsilon_{xx}m_{11} + \epsilon_{yy} m_{12} + \epsilon_{zz} m_{13}
\\= \epsilon_{xx}(m_{11} - \nu_P m_{12} - \nu_S m_{13})\\
(\Delta\rho/\rho)_{yy} = \epsilon_{xx}(m_{12} - \nu_P m_{11} - \nu_S m_{13})
\end{eqnarray}

where $\nu_P = -\epsilon_{yy}/\epsilon_{xx}$ is the Poisson's ratio of the PZT stack and $\nu_S = -\epsilon_{zz}/\epsilon_{xx}$ is the effective Poisson's ratio of the sample glued to the stack. We take the difference of the change of the resistivity of the two measurements to obtain:
\begin{eqnarray}
(\Delta\rho_{xx}- \Delta\rho_{yy})/\rho
= (\epsilon_{xx}-\epsilon_{yy})(m_{11} - m_{12})
\end{eqnarray}
The normalization factor, $\rho = \sqrt{\rho_{xx}\rho_{yy}}$, is calculated for each sweep using $\rho_{xx}$ and $\rho_{yy}$ values taken at zero anisotropic strain. If the dynamic range of the PZT is not large enough to tune through the zero anisotropic strain point the zero strain values are determined by linearly extrapolating $\rho_{xx}$ and $\rho_{yy}$ to the value at which they are equal.  Since zero anisotropic strain does not necessarily coincide with zero isotropic strain, the $B_{1g}$ response and the normalization factor are determined at a point with some residual isotropic $A_{1g}$ strain.

Expression $S16$ shows that $m_{11} - m_{12}$ is the appropriate coefficient for the $B_{1g}$ nematic susceptibility. Similarly, by cutting the crystal in a square configuration with the edges oriented at 45 degrees with respect to the tetragonal $a$-axis (so now it is experiencing $B_{2g}$ strain with respect to the crystal axes), it can be easily shown via transformation of the elastoresistivity tensor that the difference in the change of resistivities can be expressed by the following: 
\begin{eqnarray}
(\Delta\rho_{xx}- \Delta\rho_{yy})/ \rho = (\epsilon_{xx}-\epsilon_{yy})2m_{66}
\end{eqnarray}
Here $x$ and $y$ still refer to the axes along which normal strain is applied and $m_{66}$ is defined with respect to the crystal axes. Since $\epsilon_{xx}$ can be measured directly by the strain gauge glued on the back of the PZT stacks\cite{StGage} and the Poisson ratio is known \cite{HH_2013}, we can directly extract the elastoresistivity coefficients by fitting the linear slope of $(\Delta\rho_{xx} -\Delta\rho_{yy})/\rho$ as a function of strain. An example of the linear fitting can be found in Fig. S4. Further details and the associated description of the elastoresistance can be found in refs. \cite{HH_2013, Max_Sym} 

\subsection*{Geometric factor}
During an elastoresistivity measurement the sample undergoes a small physical deformation, therefore in addition to the change of resistivity that is proportional to the elastoresistivity coefficients, there is also a change of the sample dimensions (ie a geometric effect that also affects the measured resistance). In a standard four terminal resistivity measurement this can be expressed as: 
\begin{eqnarray}
\Delta\rho/\rho = \Delta A/A - \Delta L/L + \Delta R/ R
\end{eqnarray}
Where $A$ is the cross sectional area of the current path and $L$ is the separation between contacts. For the Montgomery method the equivalent expression is more complicated and is given by:
\begin{widetext}
\begin{eqnarray}
\begin{split}
\begin{aligned}
& \Delta\rho/\rho = \Delta A/A - \Delta L_{\parallel}/L_{\parallel} \\ &- \frac{\big(R_{\parallel}
\Delta R_{\perp} - R_{\perp} \Delta R_{\parallel} \big)Coth \big(\frac{ln(\frac{R_{\perp}}{R_{\parallel}}) + \sqrt{4\pi^2 + ln(\frac{R_{\perp}}{R_{\parallel}})^2}}{2\pi} \big) \big(ln(\frac{R_{\perp}}{R_{\parallel}}) +\sqrt{4\pi^2 + ln(\frac{R_{\perp}}{R_{\parallel}})^2} \big)}{2\pi R_{\parallel} R_{\perp} \sqrt{4\pi^2 + ln(\frac{R_{\perp}}{R_{\parallel}})^2}} \\ & + \frac{ 2\pi \big(-R_{\parallel}\Delta R_{\perp} + R_{\perp} \Delta R_{\parallel} \big(1 + \sqrt{4\pi^2 + ln(\frac{R_{\perp}}{R_{\parallel}})^2} \big) \big)}{2\pi R_{\parallel} R_{\perp} \sqrt{4\pi^2 + ln(\frac{R_{\perp}}{R_{\parallel}})^2}} 
\end{aligned}
\end{split}
\end{eqnarray}
\end{widetext}
where $L_{\parallel}$ is the length between the measurement leads, $R_{\parallel}$ is the resistance of the sample along the current path, and $R_{\perp}$ is the resistance of the sample perpindicular to the current flow. For an isotropic square sample ($R_{\parallel} = R_{\perp} = R$ and $\Delta R_{\parallel} = \Delta R_{\perp} = \Delta R$) this simplifies to the standard four terminal expression.

In both cases, the geometric terms $\Delta L/L$ and $\Delta A/A$ account for the change in the resistance of the crystal due solely to changes in physical dimensions and have no temperature dependence. In a simple metal, such as copper, the elastoresistivity coefficients are small, and the change of resistance is dominated by the geometric factor. However, in semiconductors or strongly correlated materials the elastoresistivity coefficients can be much larger than the geometric effect \cite{Sun_2010}. In the analysis described in the main text, the geometric factor is partially responsible for the small, temperature independent term $m_{66}^{0}$.

\subsection*{Practical considerations: Strain Transmission}

Strain is transmitted to the sample through a standard two part epoxy\cite{Epoxy} used to glue the samples on to the sidewall of the PZT stacks. Strain can be dissipated both in the epoxy layer and in the sample itself. The ability of the stack to transmit the strain to the sample therefore depends on the choice of glue and the thickness and elastic properties of the sample\cite{Marti_2013}. Previously it was found that the strain is essentially fully transmitted for large (i.e. in plane dimensions $>$2000$\mu$m) samples with thicknesses below approximately 100$\mu$m \cite{JH_2012}; measurements presented in this paper were made for samples with thicknesses $\leq$50$\mu$m. To confirm that the strain was transmitted in our samples ($\sim750\mu$m square), we prepared a series of test samples: three square BaFe$_{2}$As$_{2}$ samples of similar thickness with side lengths  $\sim300\mu$m, $\sim750\mu$m, and $\sim3000\mu$m. Strain gauges\cite{StGage} were glued both on the back side of the stack and on top of the largest test crystal (due to the large size of the strain gauge we are not able to glue the gauge on top of the smaller crystals). In Fig. S5 we show the strain transmission in the largest crystal. For temperatures below 255K, the strain transmission is $\geq80\%$ with only a weak temperature dependence. We compare the $m_{66}$ elastoresistivity coefficient calculated from the strain measured from the two (sample and PZT mounted) strain gauges in Fig. S6. The two calculations are in good agreement with each other and have Curie-Weiss critical temperatures ($T*$) values of 109.0$K$ $\pm$ 0.7$K$ and 107.7$K$ $\pm$ 0.5$K$ respectively, where uncertainties are standard errors obtained from each individual Curie-Weiss fit. The weak temperature dependence of the strain transmission seen in Fig. S5 is dominated by the much stronger Curie-Weiss temperature dependence of the elastoresistance and therefore has negligible effect on the determination of the elastoresistivity coefficient m$_{66}$ and the associated fit parameters. Details about the Curie-Weiss fit and fit parameters can be found in the main text and subsequent sections. 

Though we cannot directly measure the strain in the smaller crystals we can compare the $m_{66}$ strain response (Fig. S7). We see good agreement in the high temperature $m_{66}$ response between the $\sim3000\mu$m and $\sim750\mu$m samples, indicating a similar strain transmission in the smaller $\sim750\mu$m sample and larger $\sim3000\mu$m sample. However, the smallest sample, while looking qualitatively similar with a divergent nematic susceptibility, has a smaller response. This indicates that the strain is not fully transmitted, but we can still compare the temperature dependence of the response. By normalizing the $m_{66}$ by its value at 135$K$ ( $m_{66}(135K)$), we see that the normalized responses of the two samples are identical within uncertainties (Fig. S8). The fitted $T*$ values are 107.3$K$ $\pm$ 0.9$K$ and  109.0$K$ $\pm$ 0.7$K$ for the small and large samples respectively. The similarity in the temperature dependence indicates that the small sample behaves as if it was simply experiencing a smaller strain, which does not effect the extracted nematic critical temperature. While the majority of samples used in this paper are in the size range of full strain transmission, it is important to note that the phosphorus substituted iron arsenide samples are $\sim300\mu$m square and are in the regime of smaller strain transmission. While this affects the quantitative estimate of m$_{66}$ for these samples, it does not affect the temperature dependence. See Table S1 for a list of all sample dimensions. Comparison of the Weiss temperature extracted from these three measurements of \Ba122  samples (109.0$K$ $\pm$ 0.7$K$ for 3000$\mu m$ sample, 116$K$ $\pm$ 2.2$K$ for 750$\mu m$ sample, 107.3$K$ $\pm$ 0.9$K$ for 300$\mu m$ sample) providing an estimate of the systematic uncertainties that presumably comes from variation in sample adhesion, which for example could affect the amount of residual A$_{1g}$ strain that could potentially change the system's bare nematic critical temperature.

\subsection*{Comparison to differential longitudinal elastoresistance using two crystals}

Additional data were collected using the differential longitudinal elastoresistance technique, in which two bar like samples were used to obtain $\rho_{xx}$ and $\rho_{yy}$ separately, following the method we originally described in ref \cite{HH_2013}. In comparison to the newly developed modified Montgomery's method that only requires single square shape sample, the bar shape sample suffers from strain relaxation along the direction perpendicular to the current if the width of the sample is comparable to its thickness. Nevertheless for materials with a elastoresistance dominated by the response in nematic channel, these two techniques yield qualitatively similar results. 

To demonstrate this point, data collected by differential longitudinal technique are shown in Fig. S9 (four different optimally doped iron-based superconductors), and Fig. S10 (four dopings of \CoBa122). All the features presented in Fig. 2 and 3 in the main text including diverging nematic susceptibility for optimally doped sample, perfect Curie-Weiss temperature dependence for P doped \Ba122 , and deviation from Curie-Weiss behaviors near optimally doping for \CoBa122 are reproducible. These agreements further corroborate the robustness of our findings.

\section*{Sign and magnitude of m$_{66}$ of Ba$_{0.6}$K$_{0.4}$Fe$_2$As$_2$}
The main subject of our paper is the observation of a divergence of the differential elastoresistance that we have found for many different optimally doped compositions of Fe-based superconductors. However, beyond this main result several other features of the data are of interest in their own right and deserve additional comment. In this section we particularly emphasize two aspects of the elastoresistance of optimally doped Ba$_{0.6}$K$_{0.4}$Fe$_2$As$_2$. First; whereas our measurements, described in the main text, reveal that the sign of the $m_{66}$ elastoresistivity coefficient is \emph{negative} for Co, Ni and P substituted \Ba122 , we find that it is \emph{positive} for K-doped \Ba122. Second, the absolute magnitude of the coefficient is comparable for all four materials. Both of these observations warrant some additional commentary. 

For Co, Ni and P substituted \Ba122, the negative sign of m$_{66}$ in the tetragonal phase (for temperatures above the structural transition temperature, $T_S$) is consistent with the resistivity anisotropy that is observed for detwinned crystals in the orthorhombic state for temperatures \emph{below} $T_S$ \cite{Chu_2010,HH_2011,HH_2012}. That is, the resistivity in the tetragonal phase \emph{decreases} as the sample is stretched along the current direction. 

For \KBa122 , the in-plane resistivity anisotropy has also been measured up to $x = 0.26$ \cite{Blomberg_2013}. It was observed that for $x < 0.235$, the sign of resistivity anisotropy is the same as that of the parent compound and Co-doped \Ba122. But for $x \geq 0.235$, the sign of resistivity anisotropy was found to reverse; i.e. the resistivity was found to be \emph{higher} for the longer lattice constant direction. Our observation of a positive value for $m_{66}$ for optimal-doped \KBa122\, which has a potassium concentration $x = 0.4$, confirms this dramatic change in sign of the resistive anisotropy for sufficient hole doping. Significantly, our measurement is made in the tetragonal state, and therefore is not complicated by issues associated with the reconstruction of the Fermi surface in the antiferromagnetic state, or by concerns associated with detwinning crystals in the orthorhombic state. 

Despite the agreement in terms of the sign of the resistive anisotropy for sufficiently doped \KBa122, the relative magnitude of the anisotropy of detwinned underdoped compositions is quite different to the strain-induced elastoresistance near optimal doping. The magnitude of the resistivity anisotropy of detwinned crystals of underdoped K-substituted \Ba122 was found to be rather small \cite{Blomberg_2013}. This observation has been interpretted by some authors as evidence supporting the notion that the resistivity anisotropy derives from anisotropic scattering associated with lattice defects, which acquire an anisotropy due to the nematic fluctuations\cite{Brian_2014}. Within such a scenario, defects in the FeAs planes are believed to have a stronger effect, and hence would naturally yield a larger resistivity ansiotropy, than defects that lie away from the FeAs plane (as would be the case for K substitution), potentially accounting for the significantly lower resistivity anisotropy observed for detwinned underdoped \KBa122 relative to \CoBa122. Our measurements, however, reveal a comparable magnitude (within a factor of two) for $m_{66}$ of optimally doped Co-substituted and K-substituted \Ba122,  in contrast to the much larger difference in the resistivity anisotropy in the orthorhombic state for underdoped compositions. Clearly K substitution per se does not necessarily imply a small resistive anisotropy.  

There are three possible explanations for the small anisotropy that was previously measured in the orthorhombic state for K-substituted BaFe$_2$As$_2$. Firstly, it is possible that the resistivity anisotropy has a small magnitude simply because it changes sign as a function of composition, and therefore must go through zero for some composition. Secondly, it is possible that the FS reconstruction in the antiferromagnetic state affects the resistivity anisotropy differently for electron and hole doped cases. Finally, it is also possible that the samples used for the study of underdoped compositions were not fully detwinned, although this seems less likely given the careful characterization of the twin populations by the authors \cite{Blomberg_2013}. At present, it is not clear which of these possibilities is responsible for the small resistivity anisotropy of underdoped samples of \KBa122 in the antiferromagnetic orthorhombic state. 

Bearing this in mind, P-substitution gives another useful insight to the origin of the resistive anisotropy. Beyond the residual resistivity, another good measure of the disorder affecting transport in the Fe plane is the presence or absence of quantum oscillations. P-substituted samples exhibit quantum oscillations across the phase diagram (at least on the overdoped side\cite{Jim_QO, Shishido_2010},) indicating a much lower elastic scattering rate than the other compounds we have discussed so far, for which quantum oscillations are not observed. Significantly, $m_{66}$ of the P-substituted samples is comparably large to that of Co, Ni and K doping, despite the lower elastic scattering rate, implying that the physical origin of the dominant contribution to the resistive anisotropy is unlikely to be associated with scattering from anisotropic defects, contrary to suggestions advanced in ref. \cite{Brian_2014}, even though such effects are necessarily present. 

The observations described above serve to emphasize an important point; that elastoresistance measurements for $T > T_s$ are the most relevant data to compare with theories of the resistivity anisotropy in the nematic phase, rather than comparison with transport anisotropies in the low T Neel and/or orthorhombic state measured in samples detwinned by uni-axial stress. There are several reasons, but primarily because (a) there are no concerns about degree of detwinning (in the tetragonal phase the material is homogeneous, and homogeneously strained); (b) there are no issues with back-folding of bands and FS reconstruction; and (c) we have precise knowledge of the strain experienced by the material (which is not known for detwinned samples, for which the amount of strain depends on the temperature dependent elastic modulus). In other words, in selecting transport data to compare with theoretical predictions for resistive anisotropy, elastoresistance measurements provide the more quantitatively controlled experimental perspective.  

\section*{Zero strain resistivity versus temperature of the optimal doped Fe-based superconductors}
For reference, we show in Figure S11, the temperature-dependence of the resistivity for the same materials for which elastoresistance data are shown in the main text. The data were taken for free-standing (unstrained, and unattached) crystals. Data are plotted as R/R(300K) to eliminate uncertainty in geometric factors.

\section*{Elastoresistivity coefficients of \FeTeSe}
The "11" family of iron chacogenides (\FeTeSe) provide an interesting example of the richness of the phase diagram of iron based superconductors. At the tellurium end of the phase diagram, Fe$_{1+\delta}$Te possesses a different antiferromagnetic ground state; the antiferromagnetic order is a bicollinear order with an ordering wave-vector that is 45$^\circ$ to that in iron pnictides\cite{Bao_2009}. Consequently the spontaneous strain developed in the symmetry breaking phase is $\epsilon_{xx} - \epsilon_{yy}$ (B$_{1g}$) rather than $\epsilon_{xy}$ (B$_{2g}$). On the other hand, at the selenium end of the phase diagram, FeSe only has a structural transition that develops a spontaneous B$_{2g}$ strain $\epsilon_{xy}$ without any long range magnetic order\cite{McQueen_2009}. The optimal doped FeTe$_{0.6}$Se$_{0.4}$ we measured in this work has a $T_c = 13K$ and there is no sign of phase transitions other than the superconducting transition. 

We have measured $m_{11} - m_{12}$ (B$_{1g}$) elastoresistivity coefficient of FeTe$_{0.6}$Se$_{0.4}$, Ba$_{0.6}$K$_{0.4}$Fe$_2$As$_2$ and Ba(Fe$_{0.93}$Co$_{0.07}$)$_2$As$_2$. These measurements were taken using a differential elastoresistance technique which involves simultaneously measuring the resistivity of two orthogonal bar-like samples. Details of this technique can be found in ref. \cite{HH_2013}. The results are plotted in Fig. S12 together with the $m_{66}$ (B$_{2g}$) coefficient presented in Fig. S9. It can be clearly seen that the nematic suscptibility diverges only in the B$_{2g}$ channel rather than the B$_{1g}$ channel. 

\section*{Further details about the Curie-Weiss fit parameters}
As described in the main text, $m_{66}$ data were fit to a Curie-Weiss temperature dependence: 

\begin{equation}
2m_{66} = 2m_{66}^0 + \lambda/[a(T-T^*)]
\end{equation}  

where $T*$ gives the bare mean-field nematic critical temperature (the temperature at which nematic order would occur if there were no coupling to the crystal lattice \cite{JH_2012, HH_2013}), and $\lambda/a$ is the Curie constant. 

The fitting is performed by varying $T*$ and $2m_{66}^0$ until the standard deviation of $(2m_{66} - 2m_{66}^0)(T-T^*)$ is minimized, in which $2m_{66}$ and $T$ are experimentally measured data. This fitting method is justified by the fact that $(2m_{66} - 2m_{66}^0)(T-T^*) = \lambda/a$ is expected to be a constant as a function of temperature.

The determination of fitting temperature range is not a trivial task, there are systematic deviations from Curie-Weiss behavior in both low temperatures (due to the effect of disorder near quantum critical point) and high temperatures (due to the vanishing nematic fluctuations). In order to unbiasedly determined the high and low temperature cutoff for the fitting, we have developed a standardized procedure: A fitting is first performed over the widest possible temperature range in which the strain is still effectively transmitted, i.e. from $T_S$ (for underdoped samples) or $T_C$ (for optimal and overdoped samples) to 250$K$. The $T^*$ and $2m_{66}^0$ determined by this fitting is then used to calculate the deviation of $(2m_{66} - 2m_{66}^0)(T-T^*)$ from its mean value as a function of temperature. An example for optimally doped \CoBa122 sample is shown in the top panel of Fig. S13. There are systematic deviations in both low and high temperature ranges, which is typically observed in all the data sets. The next step is to lower the higher temperature cutoff to the deflection point indicated by the arrow in Fig. S13, then perform the fitting in the new temperature range. The deviation of $(2m_{66} - 2m_{66}^0)(T-T^*)$ is calculated again, as shown in the second panel of Fig. S12. In the case of optimally doped \CoBa122, the systematic deviations has been reduced yet they are still observable. Therefore the next step is to increase the lower temperature cutoff until the systematic deviations is smaller than the intrinsic noise level, as shown in the third panel of Fig. S13. Finally, we raise the upper temperature cutoff again until the systematic deviation appears to be greater than the intrinsic noise level (4th panel in Fig. S13).

For heavily underdoped \CoBa122, and for optimally doped BaFe$_2$(As$_{0.7}$P$_{0.3}$)$_2$ and \FeTeSe, all the systematic deviations can be eliminated at the second stage of the standardized procedure, i.e. there is no need to raise the lower temperature cutoff, and $m_{66}$ follows a perfect Curie-Weiss temperature dependence all the way down to $T_s$ or $T_c$. For optimally doped Co, Ni and K-substituted \Ba122, the data can be described by Curie-Weiss behavior only above a temperature that is higher than $T_c$. The range of temperatures over which this results in a reasonable description of the temperature-dependence of $m_{66}$ is shown by the linear fit (red line) to the inverse susceptibility $|2m_{66}-2m_{66}^0|^{-1}$ in the lower panels of Figures 2 and 3 in the main text. Four additional dopings of \CoBa122 were measured and the data were used to construct the color map phase diagram in Fig. 4. Fig. S14 shows the raw data of all eight dopings of \CoBa122 and their CW fittings. Associated fit parameters of all the data presented in the main text and supplementary materials are listed in Table S1.

The quality of fit to Curie-Weiss behavior for optimally substituted BaFe$_2$(As$_{0.7}$P$_{0.3}$)$_2$ can also be appreciated when the data are plotted on a log-log scale (Fig. S15). As can be seen, only the optimally P-substituted data (red) can be fitted linearly over the whole temperature range, with a slope $\gamma = 0.985 \pm 0.005$ (solid black line). For all other optimally-doped samples, the data exhibit a deviation from Curie-Weiss behavior. 
 
In addition, we show below the elastoresistance coefficient $2m_{66}$ of a P-substituted sample which is slightly overdoped with respect to the sample shown in the main paper. As shown in Fig. S16, for this slightly overdoped composition the elastoresistance still follows a perfect Curie-Weiss temperature dependence, but now the Weiss temperature is -11.5 K. For the approximately optimally doped composition shown in the main text, the Weiss temperature is + 11.7K. Clearly the Weiss temperature crosses zero, indicative of a nematic quantum critical point.  

\section*{Theory of the Effect of Weak Disorder on the Nematic Susceptibility}

Here we analyze the properties of an electron nematic that derive primarily from general symmetry considerations. In particular, we discuss how the presence of quenched disorder affects the behavior of the nematic susceptibility upon approach to a nematic thermal or quantum critical point. Admittedly, many aspects of the measurements we have reported cannot be understood in this way as they  depend on material-specific details and require an explicit understanding of the  processes that contribute to the transport anisotropy, specifics of the particular type of quenched disorder involved, and of course an understanding of the underlying thermodynamic mechanism of nematicity. Nevertheless, this simple analysis serves to offer intuition for the effects of disorder in nematic systems. It also reveals effects that have a remarkable similarity to the observed deviations from mean field Curie-Weiss behaviors of the elastoresistance for disordered optimally doped compounds presented in the main text.

We start by defining the local nematic order $\phi(\vec r)$,  which is even under inversion but odd under rotation by $\pi/2$ or reflection through the appropriate mirror plane.  The uniform thermal average of $\phi$  is the nematic order parameter, $N= \Omega^{-1}\sum_{\vec r} \langle \phi(\vec r)\rangle$, where $\Omega =\sum_{\vec r}$ is proportional to the volume.  In computing the susceptibility, we will consider the effect of a symmetry breaking field, $h$, which is conjugate to $N$, and which thus represents the appropriate element of the strain tensor -- we define the coupling so that positive $h$ favors positive $N$.  Finally, the quenched disorder is represented by a random field $b(\vec r)$ which couples in the same way as $h$ to the nematic order parameter, but is assumed to have average strength zero and to be short-range correlated, $\overline {b(\vec r)}=0$ and $\overline {b(\vec r)b(\vec r^\prime)} = \sigma^2\delta{\vec r,\vec r^\prime}$, so $\sigma$ is the relevant measure of the strength of the disorder.

{\bf Weak disorder}:  If the disorder  is weak and the system is not too close to criticality, the effect of disorder on the nematic susceptibility can be computed perturbatively in the disorder strength, which yields a result of the form

\begin{eqnarray}
N = \chi(T,\sigma)h - \chi_3(T,\sigma)h^3
\end{eqnarray}

\begin{eqnarray}
\chi^{-1}(T,\sigma)  =\chi^{-1}(T) + \sigma^2\Gamma(T) + \ldots ,
\end{eqnarray}
where $\chi(T,0)\equiv\chi(T)$, $\chi_3(T,0)\equiv \chi_3(T)$, and $\Gamma(T)$ are all related to equilibrium properties of the disorder free system (i.e. $\sigma = 0$, for which formally exact expressions can be readily obtained in terms of various correlation functions of $\phi$), and  $\ldots$ refers to higher order terms in powers of h and $\sigma$.  
 
We have defined signs in these equations such that under typical circumstances $\chi$, $\chi_3$, and $\Gamma$ are all positive quantities. 
To obtain explicit expressions, we can consider the problem in the context of a Landau-Ginzberg free energy of the form
\begin{eqnarray}
\begin{aligned}
F[\phi]=&\frac 1 2 \sum_{\vec r,\vec r^\prime} \phi(\vec r) G^{-1}(\vec r-\vec r^\prime)\phi_{\vec r^\prime} \\& + \frac u 4 \sum_{\vec r} \phi^4_{\vec r} - \sum_{\vec r} [h+b_{\vec r}]\phi(\vec r) ,
\end{aligned}
\end{eqnarray}
which we treat as a mean-field functional which we solve by minimizing $F$ with respect to $\phi$.  In the disordered phase (where $N\to 0$ as $h\to 0$), this gives rise to the expressions
\begin{eqnarray}
\begin{aligned}
&\chi(T) = g_1, \\ &\Gamma(T) = 3u g_2\sim u [g_1]^{(4-d)/2}, \\ &\chi_3 = u[g_1]^4
\end{aligned}
\end{eqnarray}
where
\begin{eqnarray}
g_n \equiv \sum_{\vec r}\big[G(\vec r)\big]^n
\end{eqnarray}
and where, for the usual reasons, $g_1 \sim (T-T_c)^{-1}$.

This simple analysis already captures some qualitative features of the  observed phenomena:  a) It shows that generically  the effect of even weak disorder grows stronger upon approach to criticality, and conversely that at temperatures well away from criticality the effects can be negligible as far as the initial growth of critical correlations is concerned.  b) Moreover, it shows that generically the first effect of disorder is to slow the growth of the susceptibility relative to its expected behavior in the absence of disorder.  However, from this perturbative perspective, there is little reason to expect larger effects upon approach to a ($T_c=0$) quantum critical point than upon approach to a classical ($T_c >0$) critical point.

{\bf Non-perturbative effects and quantum Griffith phases}:  However, it is well known that disorder -- even weak disorder -- can produce non-perturbative corrections to the thermodynamic properties of a system, and even change the analytic structure of the free energy by producing so-called Griffith singularities.  For classical finite temperature systems, these Griffith singularities tend to be unobservably weak.  However, for quantum systems with a discrete order parameter these effects are highly amplified producing a new set of phenomena known as quantum Griffith phases\cite{DSFisher_1992}. In particular, based on symmetry, we would tend to associate the universal critical phenomena associated with a continuous  quantum phase transition to a nematic ground state with those of the transverse field Ising model.  

The nature of the quantum Griffith phenomena associated with the transverse field Ising model in the presence of random field disorder have been rigorously established in 1D\cite{DSFisher_1992}, but scaling arguments which are successful in accounting for these results produce compelling results for higher dimensional problems as well\cite{Motrunich_2000}.  Among other things, these arguments imply that for a range of transverse fields $h_\perp$ greater than a critical value, $h_{\perp,c}$, the susceptibility diverges in the limit $T\to 0$ as
\begin{eqnarray}
\chi(T) \sim [1/T]^{1 -\theta}   \ {\rm with} \  \theta \sim (h_\perp - h_{\perp,c})
\end{eqnarray}
 even though the ground-state does not exhibit nematic order.

At core, the enormous difference between classical and quantum Griffith effects is illustrative of a more general tendency for the effects of disorder to be more dramatic on quantum critical phenomena than on their classical cousins.  In classical statistical mechanics, a sum is performed over all position dependent field configurations with an appropriate Boltzman weight, and quenched disorder implies the existence of  point-like defects that locally favor one particular value of the field.  In quantum statistical mechanics, the same sum is performed over field configurations that are a function of position and imaginary time, and quenched disorder implies the existence of line defects that are infinite in extent in the imaginary time direction.  Thus, even rare impurities can have a large qualitative effect on the state of the system.

It seems to us that the fact that the observed effects of disorder on the nematic susceptibility appear strongest where the extrapolated nematic ordering temperature is close to T=0 reflects the enhanced sensitivity of quantum critical phenomena to disorder.

{\bf Non-equilibrium effects}:  Finally, it is important to note that non-equilibrium effects are unavoidable in any problem related to the random-field Ising model.  The characteristic exponential slowing of dynamics upon approach to the critical temperature implies that -- necessarily --  at a temperature strictly larger than $T_c$, on any accessible laboratory time scale the system will fail to achieve thermal equilibrium.  At all lower temperatures, glassy physics involving hysteresis, and noise must be expected\cite{Carlson_2012}.  The existence or not of ``aging'' effects reflecting this physics has not been explored in the present set of experiments.

\begin{widetext}
\begin{table}[ht]
\centering 
\begin{tabular}{|c|c|c|c|c|c|}
\hline 
Materials & Fit  & $2m_{66}^0$ & $\lambda/a_0$(K) & $T^*$(K) & Sample \\
&  range(K) &  &  &  & dimension ($\mu$m)\\ 
\hline 
BaFe$_2$As$_2$ Large & 135-250 & 7.7$\pm$0.3 & -1540$\pm$13 & 109.0$\pm$0.7 & 3170$\times$2920$\times$50 \\ 
\hline
BaFe$_2$As$_2$ Medium & 135-250 & 7.3$\pm$0.8 & -1272$\pm$34 & 116$\pm$2.2 & 760$\times$750$\times$40 \\ 
\hline 
BaFe$_2$As$_2$ Small & 140-230 & 5.4$\pm$0.3 & -948$\pm$13 & 107.3$\pm$0.9 & 280$\times$300$\times$20 \\ 
\hline 
Ba(Fe$_{0.975}$Co$_{0.025}$)$_2$As$_2$ & 100-205 & 14.5$\pm$0.8 & -2706$\pm$32 & 77$\pm$0.8 & 730$\times$700$\times$30 \\ 
\hline 
Ba(Fe$_{0.953}$Co$_{0.047}$)$_2$As$_2$ & 90-200 & 9.4$\pm$0.4 & -2004$\pm$22 & 59.9$\pm$0.9 & 530$\times$580$\times$10\\ 
\hline 
Ba(Fe$_{0.945}$Co$_{0.055}$)$_2$As$_2$ & 70-180 & 25.5$\pm$0.7 & -5370$\pm$44 & 26.4$\pm$0.7 & 580$\times$540$\times$25\\ 
\hline 
Ba(Fe$_{0.94}$Co$_{0.06}$)$_2$As$_2$ & 50-197 & 27.5$\pm$1.9 & -6812$\pm$158 & -1.4$\pm$2.7 & 730$\times$760$\times$10\\ 
\hline
Ba(Fe$_{0.93}$Co$_{0.07}$)$_2$As$_2$ & 55-200 & 13.7$\pm$0.5 & -3735$\pm$40 & 1.8$\pm$1.1 & 700$\times$725$\times$10\\
\hline 
Ba(Fe$_{0.92}$Co$_{0.08}$)$_2$As$_2$ & 50-200 & 20.6$\pm$0.6 & -5758$\pm$60 & -26.1$\pm$1.4 & 600$\times$620$\times$20\\
\hline
Ba(Fe$_{0.9}$Co$_{0.1}$)$_2$As$_2$ & 38-200 & 22.8$\pm$0.6 & -8773$\pm$99 & -94.4$\pm$2.1 & 2950$\times$2630$\times$40\\ 
\hline 
BaFe$_2$(As$_{0.68}$P$_{0.32}$)$_2$ & 30-250 & 7$\pm$1.9 & -1896$\pm$58 & 11.7$\pm$3.1 & 340$\times$320$\times$30\\ 
\hline
BaFe$_2$(As$_{0.64}$P$_{0.36}$)$_2$ & 30-200 & 6.6$\pm$1.3 & -2346$\pm$48 & -11.5$\pm$2.3 & 280$\times$300$\times$10\\ 
\hline 
Ba(Fe$_{0.955}$Ni$_{0.045}$)$_2$As$_2$ & 60-210 & 12.1$\pm$0.6 & -3487$\pm$44 & 2.3$\pm$1.3 & 750$\times$700$\times$15\\ 
\hline 
Ba$_{0.6}$K$_{0.4}$Fe$_2$As$_2$ & 90-225 & -4.0$\pm$0.3 & 820$\pm$19 & 46.1$\pm$2.4 & 750$\times$760$\times$25\\ 
\hline 
FeTe$_{0.6}$Se$_{0.4}$ & 13-95 & -10.8$\pm$0.3 & -1400$\pm$17 & -11$\pm$0.7 & 930$\times$1100$\times$30\\ 
\hline 
\end{tabular}
\end{table}

\noindent \textbf{Table S1}:Fit parameters from the fit of $2m_{66}$ based on $2m_{66} = 2m_{66}^0 + \lambda/[a(T-T^*)]$ for all the compositions shown in Figures in the main text and supplementary material. 
\clearpage

\begin{figure}
\includegraphics[width=15cm]{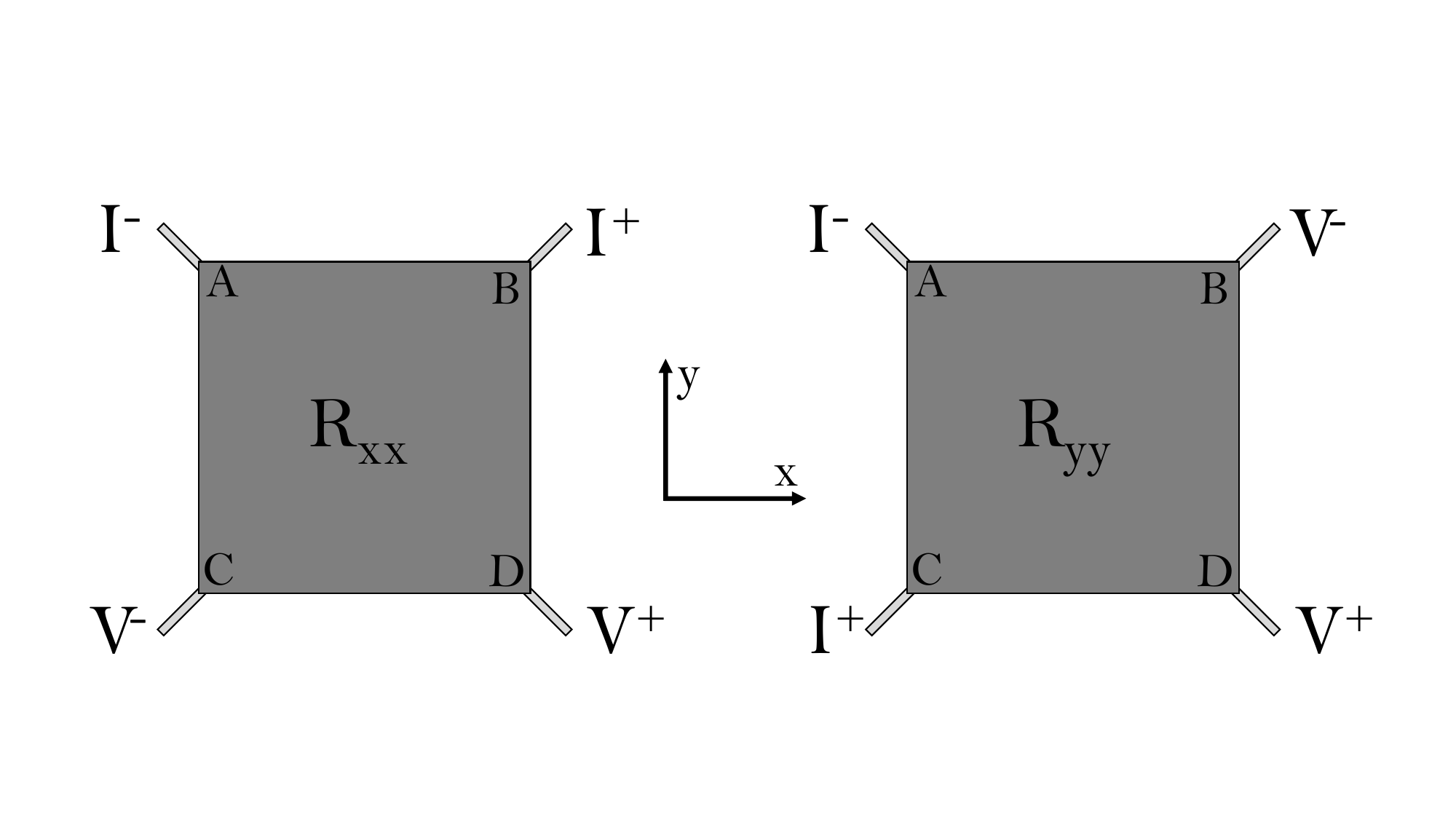}\label{FigS1}
\end{figure}

\noindent \textbf{Figure S1}:A schematic diagram illustrating the current and voltage configurations for measuring $R_{xx}$ and $R_{yy}$ using the modified Montgomery technique.  A common ground point (A) is used for both configurations. The $y$ Cartesian axis is aligned along the poling direction of the PZT stack.
\clearpage

\begin{figure}
\includegraphics[width=15cm]{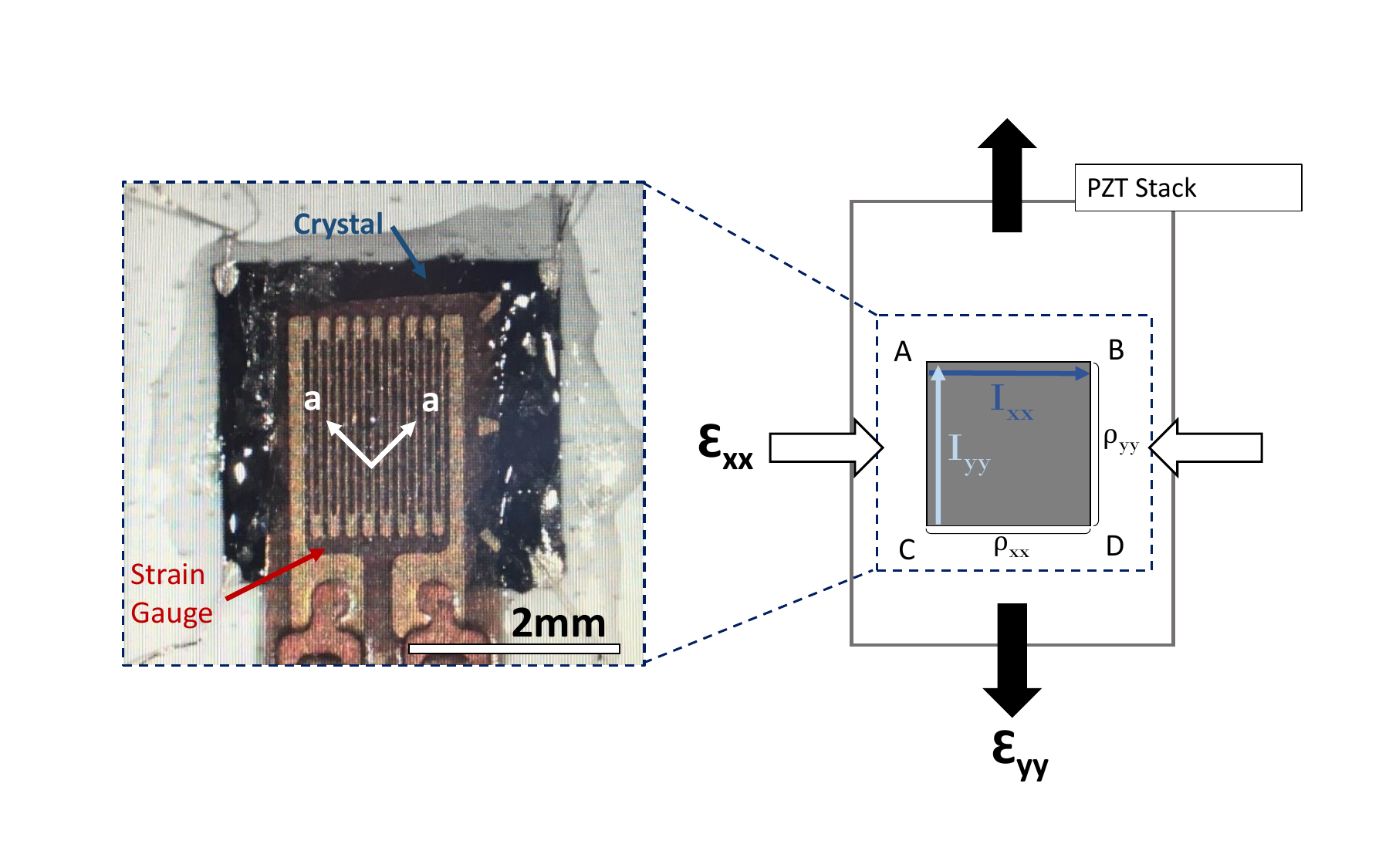}\label{FigS2}
\end{figure}

\noindent \textbf{Figure S2}: A photograph and schematic diagram illustrating measurement of $\Delta \rho_{xx}$ and $\Delta \rho_{yy}$ elastoresistance for the $B_{2g}$ symmetry channel of the parent BaFe$_2$As$_2$ compound. 
\clearpage

\begin{figure}
\includegraphics[width=11cm]{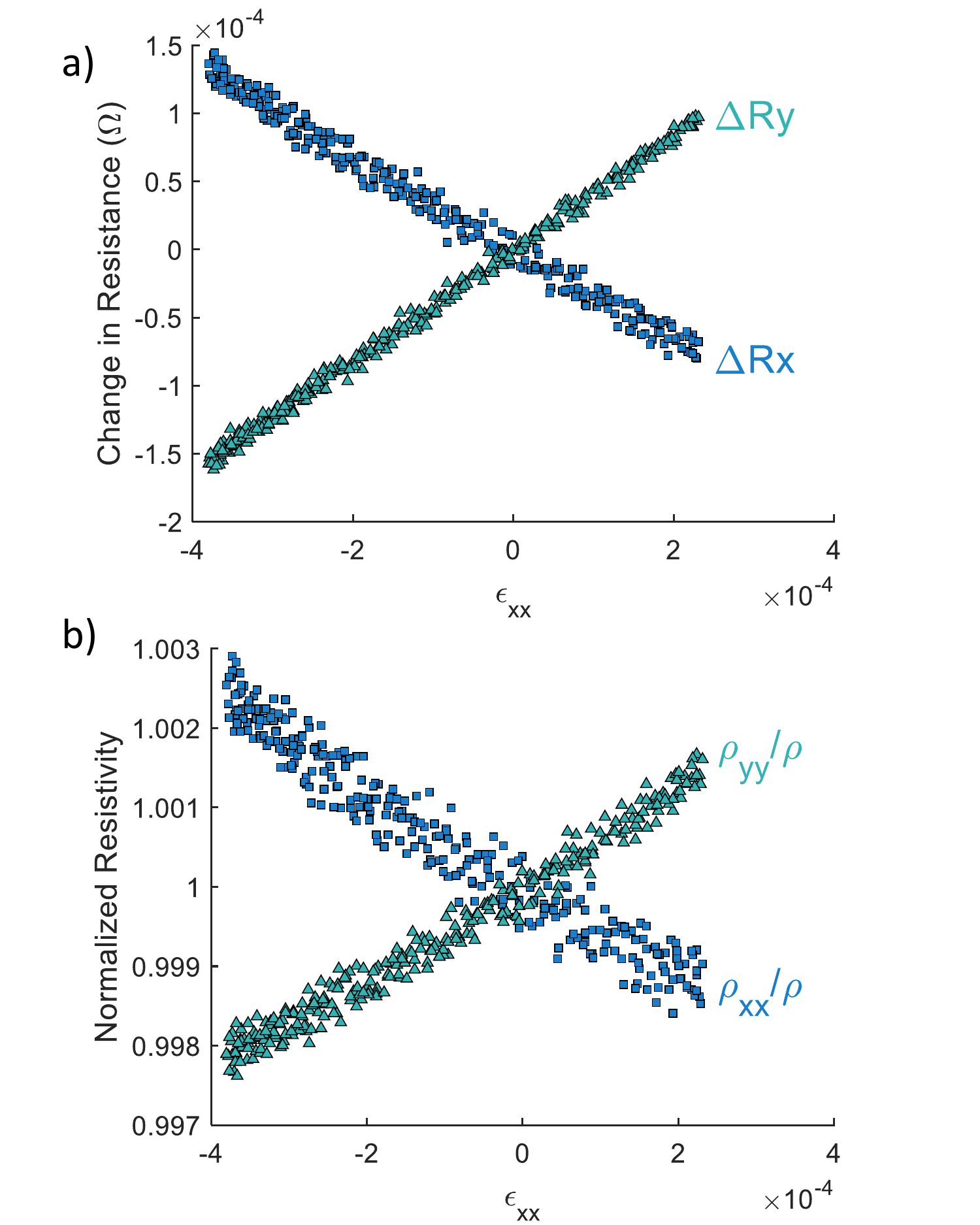}\label{FigS3}
\end{figure}

\noindent \textbf{Figure S3}:(a) A representative data set showing the measured change in resistance as a function of $\epsilon_{xx}$ strain. This data set was taken on optimally doped Ba(Fe$_{1-x}$Co$_{x}$)$_{2}$As$_{2}$ (x = 0.07) at 240K. The zero $\epsilon_{xx}$ strain point is set at zero anisotropic strain, see text for details. (b) The normalized resistivity calculated from the raw resistance data shown in (a) using a modified Montgomery method. The normalization constant $\rho$ is calculated at zero anisotropic strain as described by the main text.  
\clearpage

\begin{figure}
\includegraphics[width=15cm]{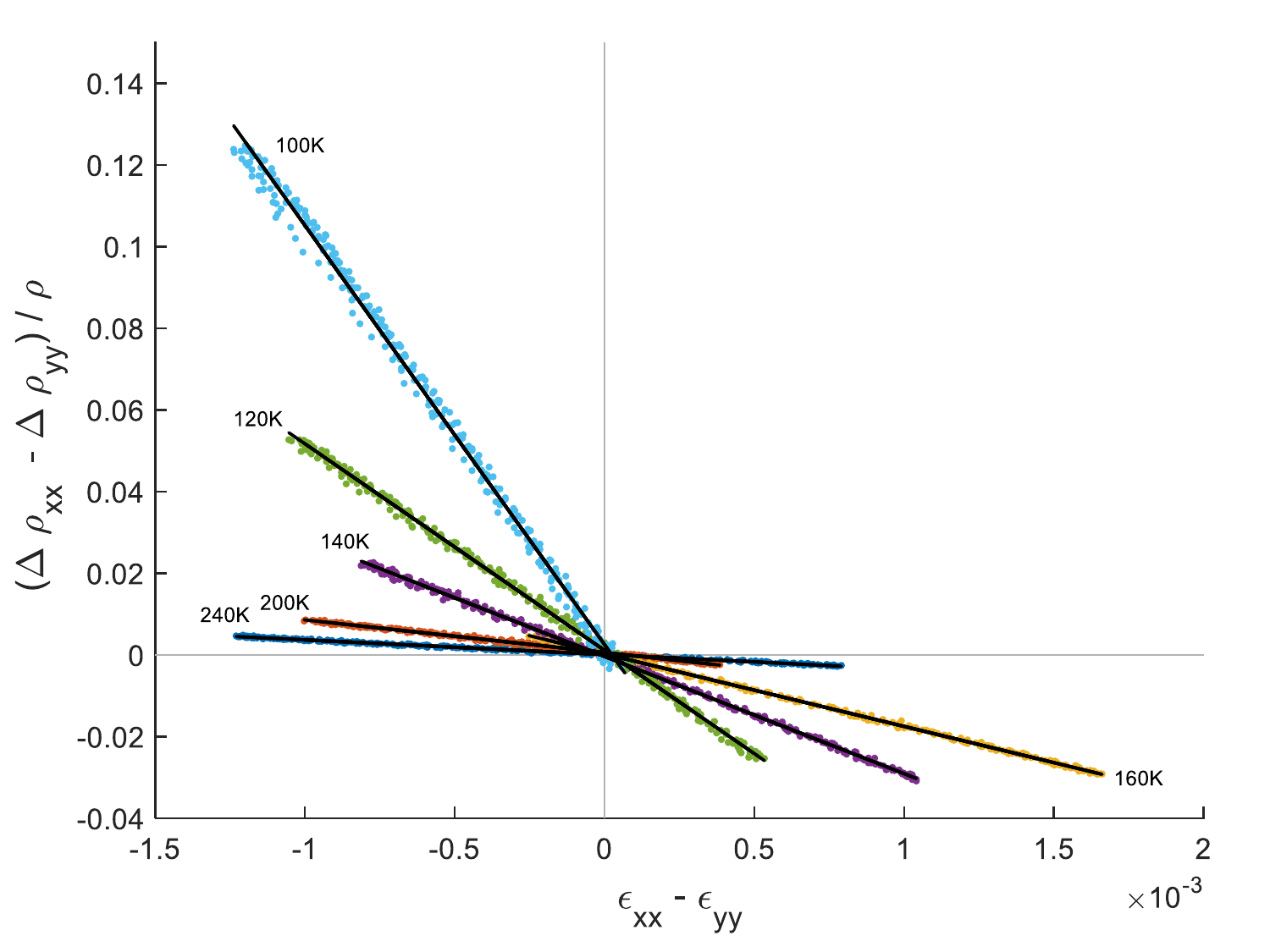}\label{FigS4}
\end{figure}

\noindent \textbf{Figure S4}: $(\Delta \rho_{xx}-\Delta \rho_{yy})/\rho$ as a function of anisotropic strain $\epsilon_{xx} - \epsilon_{yy}$ at several temperatures for the optimally doped Ba(Fe$_{1-x}$Co$_{x}$)$_{2}$As$_{2}$ (x = 0.07). The normalization constant $\rho$ is calculated at zero B$_{2g}$ strain as described by the main text. Black lines show linear fits for each temperature, from which the elatoresistivity coefficient $m_{66}$ is extracted. Strains are positive for extension, and negative for compression. 
\clearpage

\begin{figure}
\includegraphics[width=15cm]{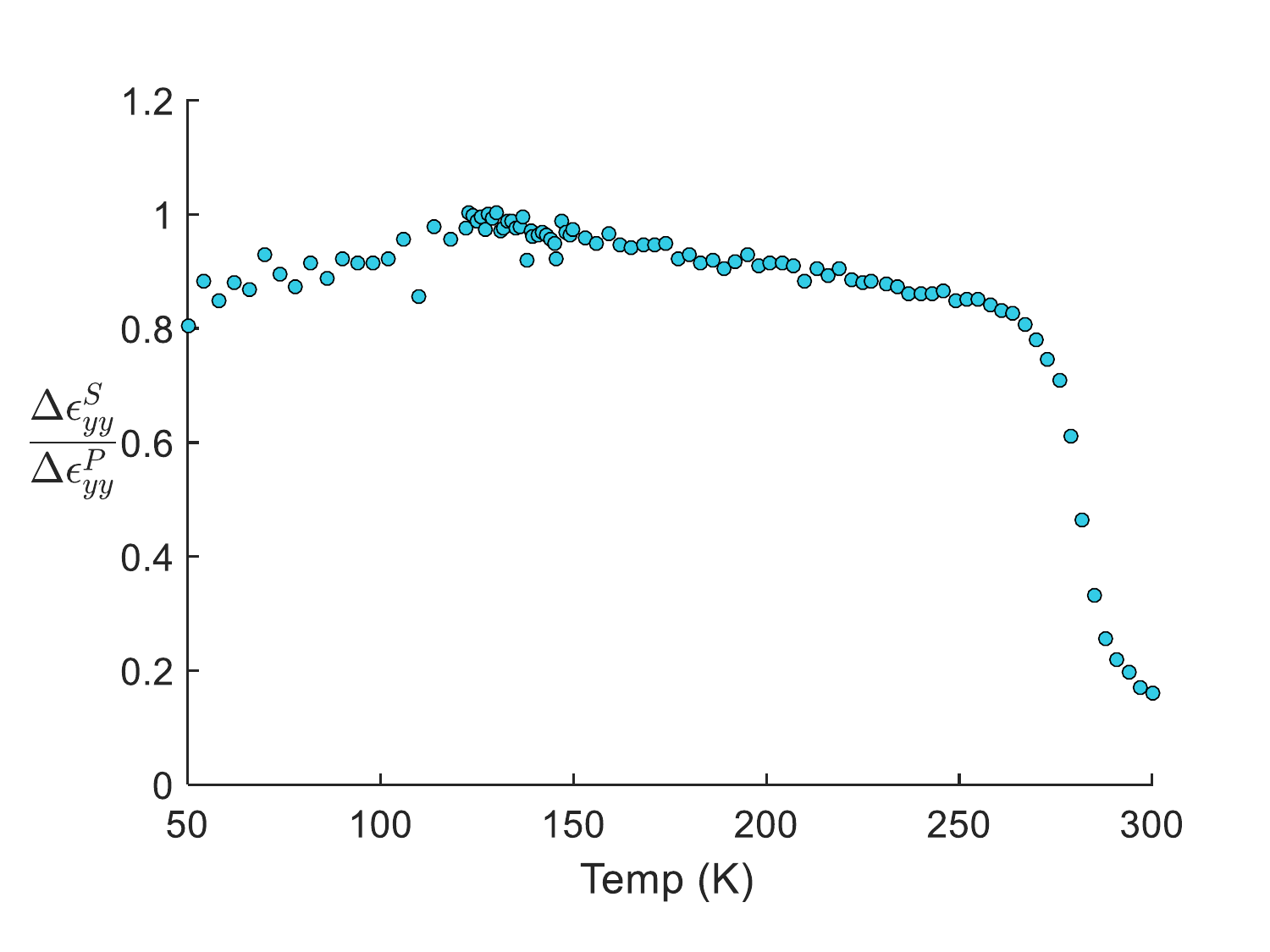}\label{FigS5}
\end{figure}
\noindent \textbf{Figure S5}: Temperature dependence of strain transmission in the large BaFe$_2$As$_2$ sample. $\Delta\epsilon_{yy}^S$ and $\Delta\epsilon_{yy}^P$ are the range of $\epsilon_{yy}$ strains experienced from a full voltage sweep applied to the PZT by the strain gauges mounted on the sample surface and PZT surface respectively. The ratio $\frac{\Delta\epsilon_{yy}^S}{\Delta\epsilon_{yy}^P}$ is the fraction of strain transmitted through the sample, for perfect transmission this ratio is 1 and for no transmission the ratio is 0. Above 255K there is poor strain transmission which we tentatively attribute to glue softening and below 255K there is good ($\geq 80\%$) strain transmission that shows a small temperature dependence. 

\clearpage

\begin{figure}
\includegraphics[width=15cm]{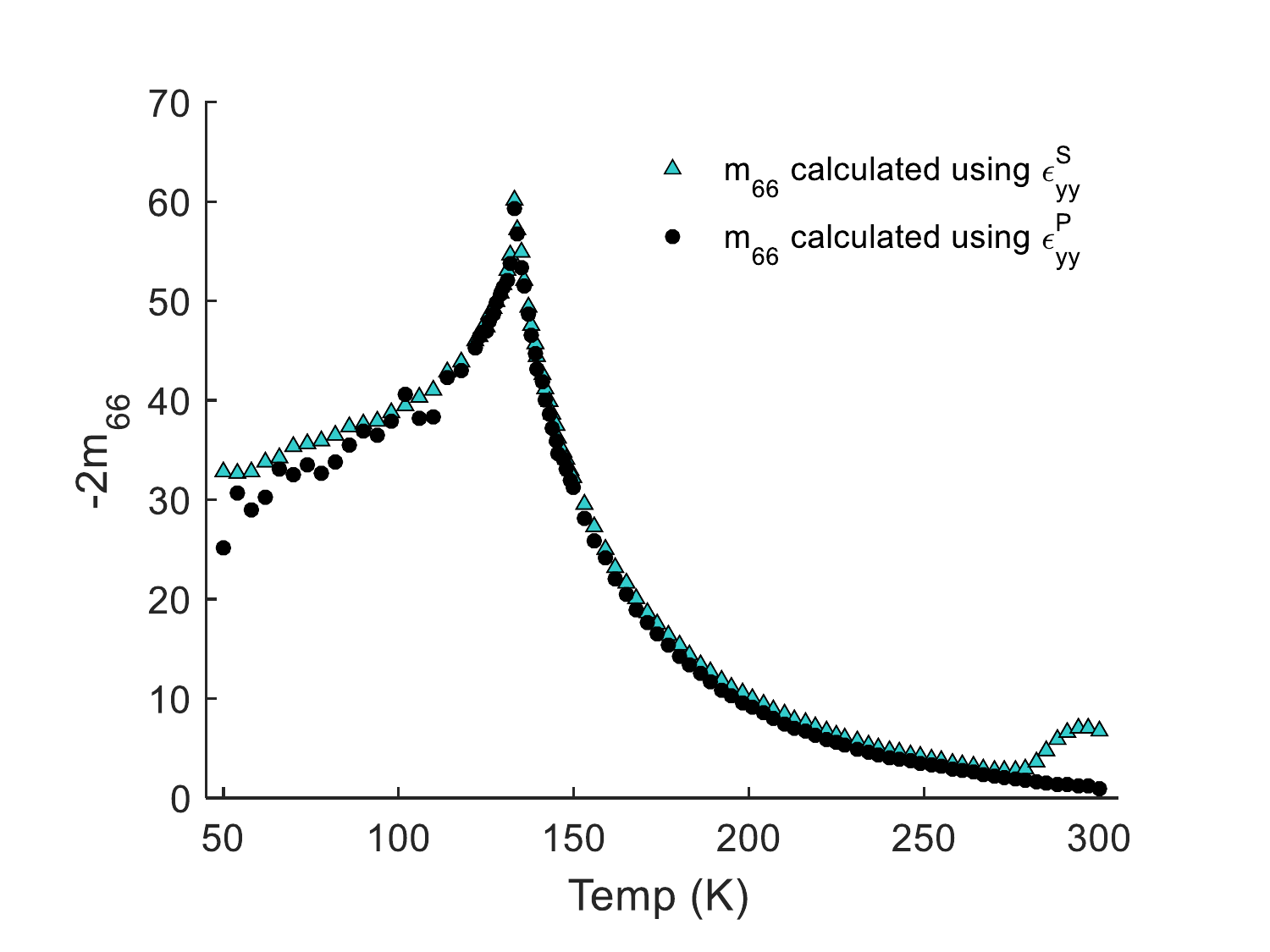}\label{FigS6}
\end{figure}
\noindent \textbf{Figure S6}: A comparison between the -2m$_{66}$ response of the large BaFe$_2$As$_2$ sample calculated from the sample and PZT mounted strain gauge. The two curves are in good agreement with each other. Curie-Weiss fitting gives $T*$ values of 109.0$K$ $\pm$ 0.7$K$ and 107.7$K$ $\pm$ 0.5$K$ respectively, indicating that the weak temperature dependence of the strain transmission below 255K does not affect the measured elastoresistivity coefficients.  
\clearpage

\begin{figure}
\includegraphics[width=15cm]{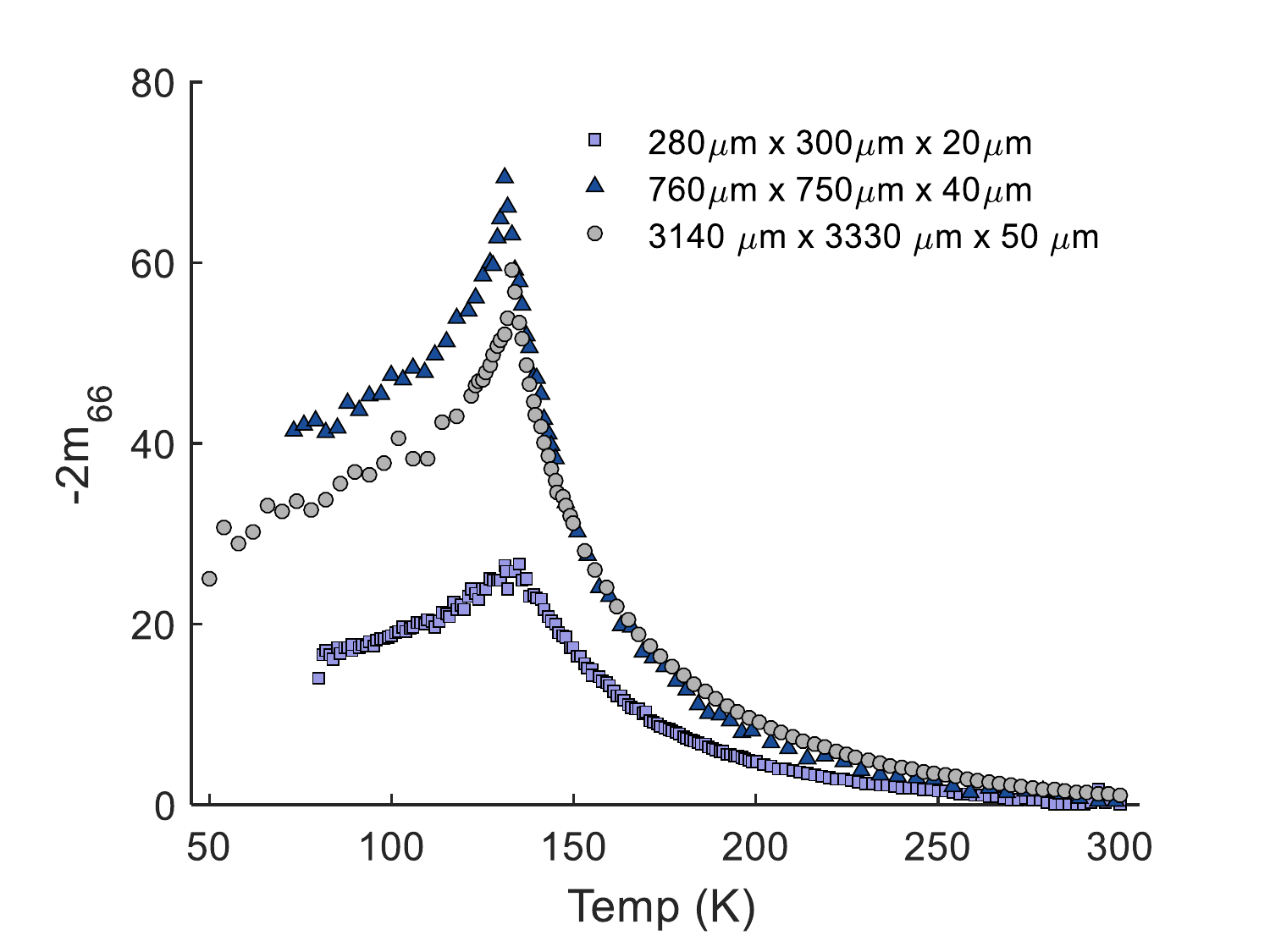}\label{FigS7}
\end{figure}
\noindent \textbf{Figure S7}: Temperature dependence of the m$_{66}$ elastoresistivity coefficient for three different sizes of undoped BaFe$_{2}$As$_{2}$ samples. The small sample  ($280\mu$m x $300\mu$m x $20\mu$m) has a smaller response than the medium ($760\mu$m x $750\mu$m x $40\mu$m) and large ($3140\mu$m x $3330\mu$m x $50\mu$m) samples. This implies equal strain transmission in the medium and large samples while a smaller strain is achieved for the small sample. 
\clearpage

\begin{figure}
\includegraphics[width=15cm]{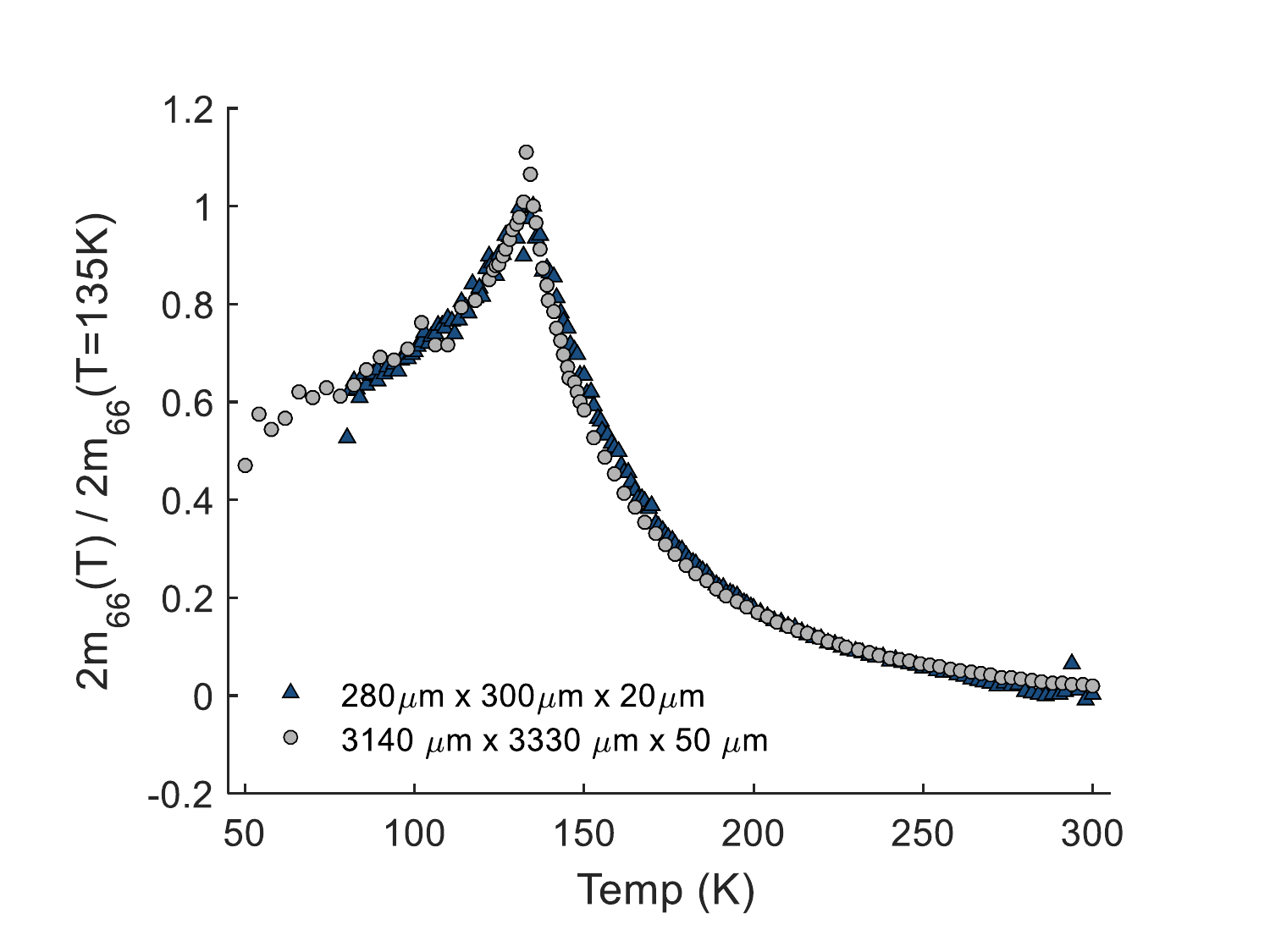}\label{FigS8}
\end{figure}
\noindent \textbf{Figure S8}: Comparison of the -2m$_{66}$ elastoresistivity coefficient normalized by the value at 135$K$ for the large ($3140\mu$m x $3330\mu$m x $50\mu$m) and small  ($280\mu$m x $300\mu$m x $20\mu$m) samples. The two curves are in good agreement with each other and show a similar temperature dependence with $T*$ values of 109.0$K$ $\pm$ 0.7$K$ and 107.7$K$ $\pm$ 0.5$K$ respectively.
\clearpage

\begin{figure}
\includegraphics[width=15cm]{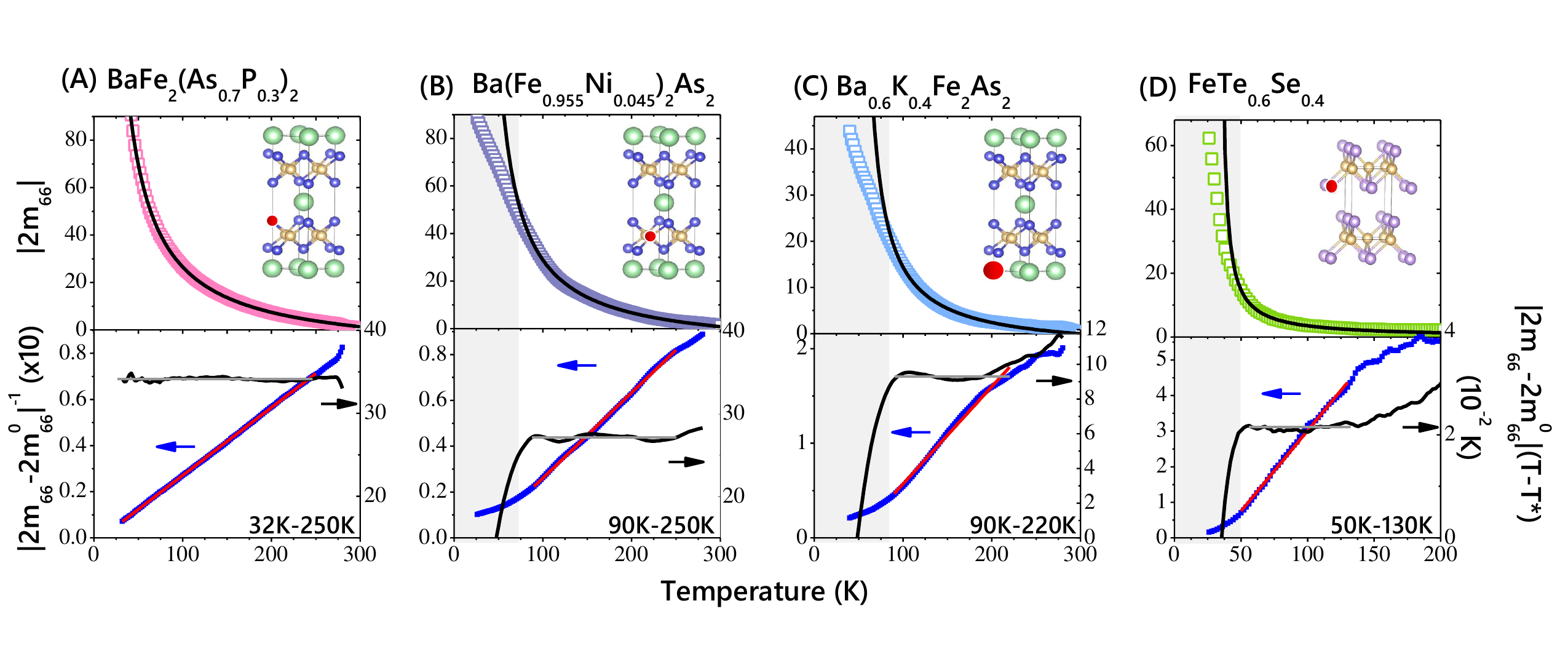}\label{FigS9}
\end{figure}
\noindent \textbf{Figure S9}: Divergence of the $B_{2g}$ elastoresistance $2m_{66}$ of several different families of optimally doped Fe pnictide and chalcogenide superconductors measured by the differential longitudinal elastoresistance technique using two separate crystals for each measurement. (A) optimally doped BaFe$_2$(As$_{0.7}$P$_{0.3}$)$_2$ (isovalent substitution),  (B) optimally doped Ba(Fe$_{0.955}$Ni$_{0.045}$)$_2$As$_2$ (electron doped), (C) optimally doped Ba$_{0.58}$K$_{0.42}$Fe$_2$As$_2$ (hole doped), and (D) optimally doped FeTe$_{0.58}$Se$_{0.42}$. Insets indicate the dopant site (red) in the respective unit cells of each material. Upper panels show $|2m_{66}|$, whereas lower panels show $|(2m_{66}-2m_{66}^0)|^{-1}$ (left axes of lower panels, blue symbols) and $|(2m_{66}-2m_{66}^0)|(T-T^*)$ (right axes of lower panels, black curves). Black(upper panels) and red(low panels) lines shows fits to Curie-Weiss behavior of $m_{66}$ and $|(2m_{66}-2m_{66}^0)|^{-1}$ respectively. Grey horizontal lines (low panels) shows the average values of  $|(2m_{66}-2m_{66}^0)|(T-T^*)$ in the fitting temperature range. Regions of deviation from Curie-Weiss behavior in (B) and (C) are indicated by gray shaded regions. For (A) and (B), $2m_{66}$ is negative. For (C) and (D), $2m_{66}$ is positive. 
\clearpage

\begin{figure}
\includegraphics[width=15cm]{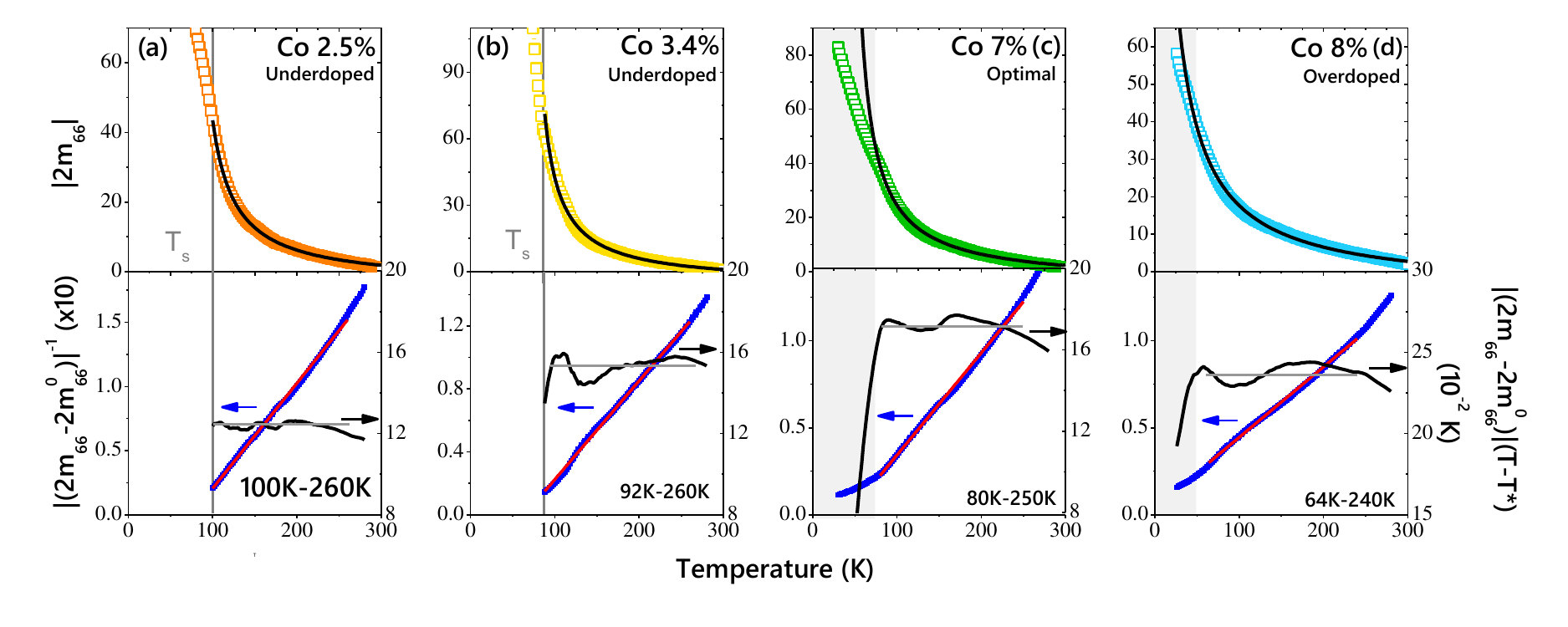}\label{FigS10}
\end{figure}
\noindent \textbf{Figure S10}: Variation of the $B_{2g}$ elastoresistance of Ba(Fe$_{1-x}$Co$_{x}$)$_2$As$_2$ for four representative compositions measured by the differential longitudinal elastoresistance technique using two separate crystals for each measurement. (A,B) underdoped compositions Ba(Fe$_{0.975}$Co$_{0.025}$)$_2$As$_2$ and Ba(Fe$_{0.953}$Co$_{0.047}$)$_2$As$_2$; (C) optimal-doped Ba(Fe$_{0.93}$Co$_{0.07}$)$_2$As$_2$; and (D) overdoped Ba(Fe$_{0.92}$Co$_{0.08}$)$_2$As$_2$. The heavily underdoped compositions is very well described by a Curie-Weiss temperature dependence over the entire temperature range(black lines in upper panels). For the compositions near optimally-doped, $2m_{66}$ can be well fit by a Curie-Weiss T-dependence for temperatures at high temperatures. Below a characteristic temperature scale(different for each compositions, indicated by shade gray region), a strong downward deviation from Curie Weiss behavior is observed, also seen in the inverse susceptibility  (upward curvature) and in $|(2m_{66}-2m_{66}^0)|(T-T^*) \propto \chi_N(T-T^*)$ (strong downturn), which are shown in the lower panels. The deviation from Curie-Weiss behavior is the strongest at the optimal doping, and diminishes on either side of the phase diagram. 
\clearpage

\begin{figure}
\includegraphics[width=17cm]{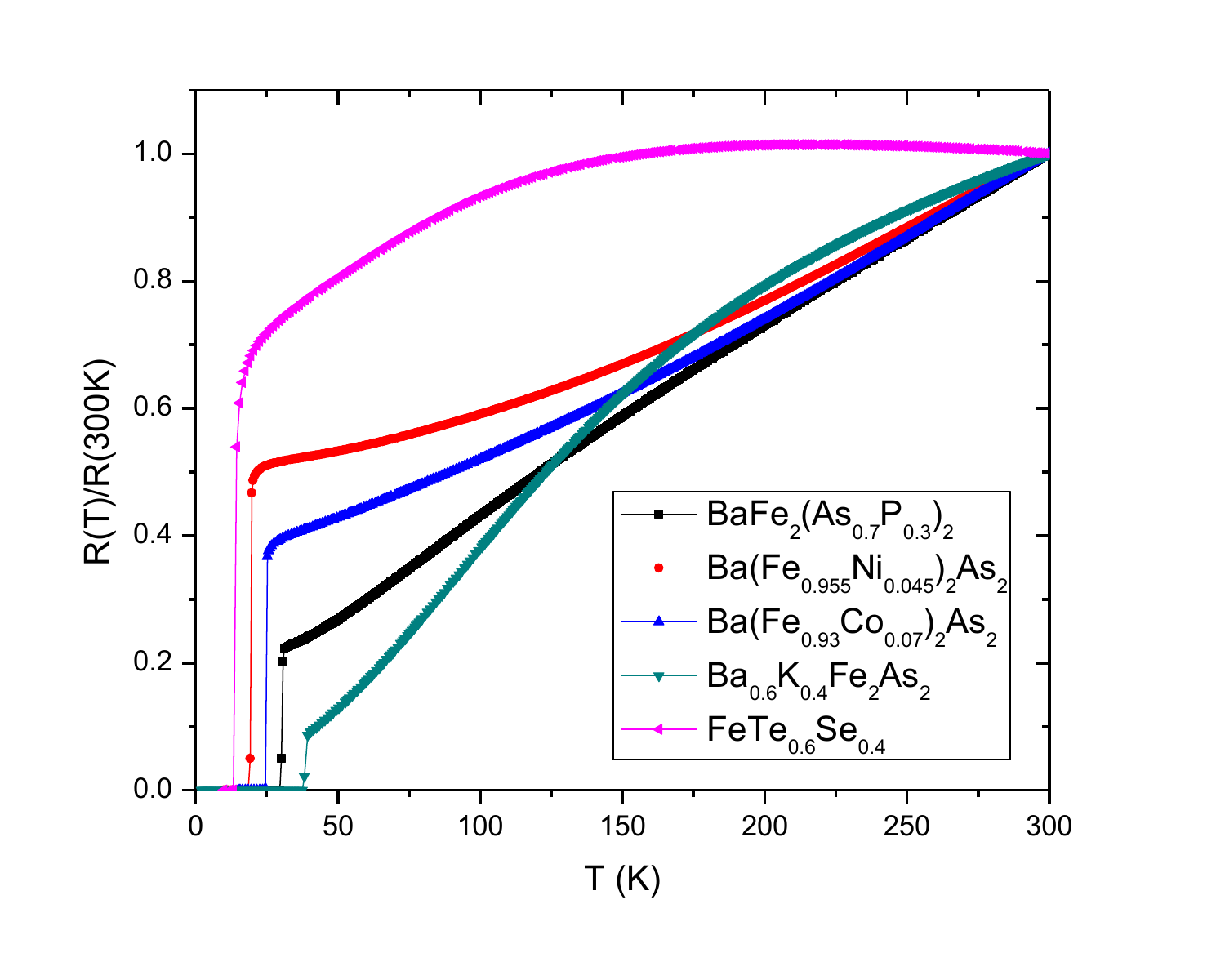}\label{FigS11}
\end{figure}
\noindent \textbf{Figure S11}: Temperature-dependence of the resistivity for the same materials for which elastoresistance data are shown in the main text. The data were taken for free-standing (unstrained, and unattached) crystals. Data are plotted as R/R(300K) to eliminate uncertainty in geometric factors.

\clearpage

\begin{figure}\includegraphics[width=17cm]{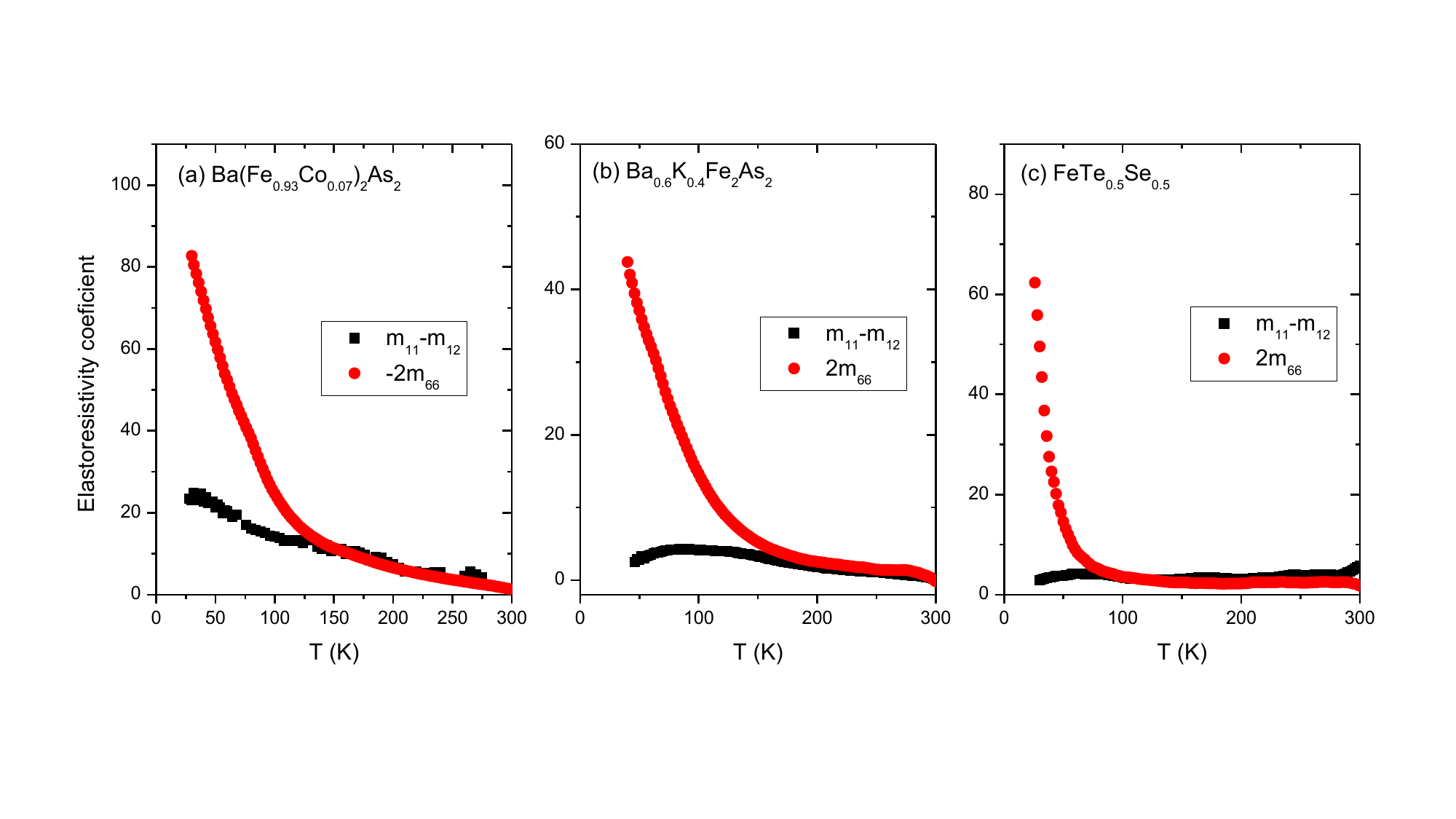}\label{FigS12}
\end{figure}
\noindent \textbf{Figure S12}: Temperature dependence of the elastoresistivity coefficients ($m_{11}-m_{12}$) and $2m_{66}$ of (a) Ba(Fe$_{0.93}$Co$_{0.07}$)$_2$As$_2$ (b) Ba$_{0.6}$K$_{0.4}$Fe$_2$As$_2$ and (c) FeTe$_{0.6}$Se$_{0.4}$. The data clearly show a divergence in the $B_{2g}$ channel (i.e. 2$m_{66}$ diverges) in all three iron based superconductors. In particular the elastoresistivity coefficients of FeTe$_{0.6}$Se$_{0.4}$ are similar to those of the Fe-arsenides, despite the fact that FeTe develops a spontaneous strain in the $B_{1g}$ symmetry channel at the structural phase transition. The data taken for this figure were from a differential elastoresistivity technique, see main text for details.

\clearpage

\begin{figure}\includegraphics[width=10cm]{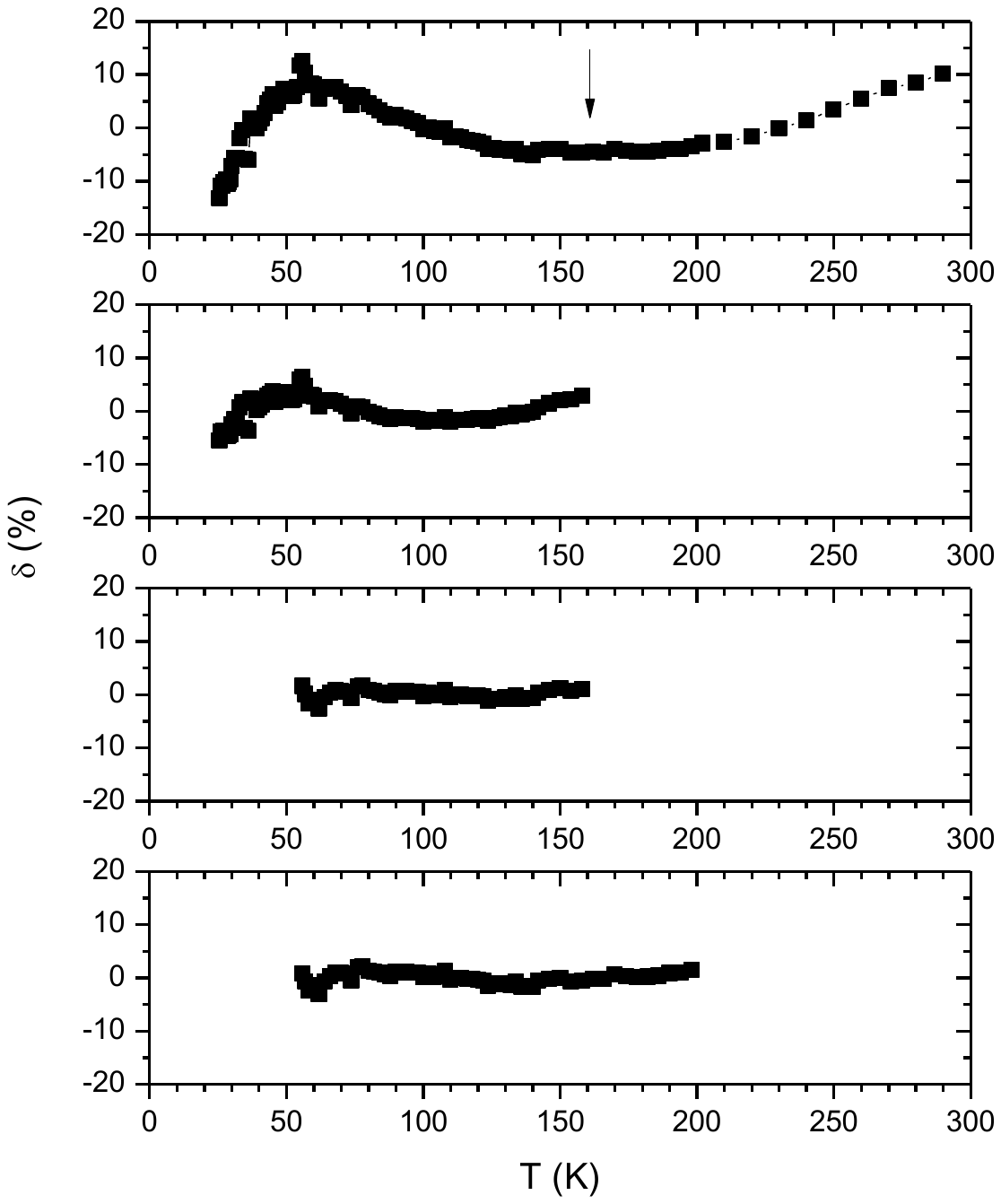}\label{FigS13}
\end{figure}
\noindent \textbf{Figure S13}: Temperature dependence of the normalized deviation $\delta$ of $\lambda/a$ for Ba(Fe$_{0.093}$Co$_{0.07}$)$_2$As$_2$. $\delta$ is calculated by taking the difference of $(2m_{66} - 2m_{66}^0)(T-T^*)$ from its mean and then normalized by the mean based on the fitted $2m_{66}^0$ and $T^*$. The fitting ranges from top to bottom are: 25 - 300K, 25 - 160K, 55 - 160K, 55 - 205K. Arrow in the top panel indicated the deflection point of high temperature deviations.

\clearpage

\begin{figure}\includegraphics[width=12cm]{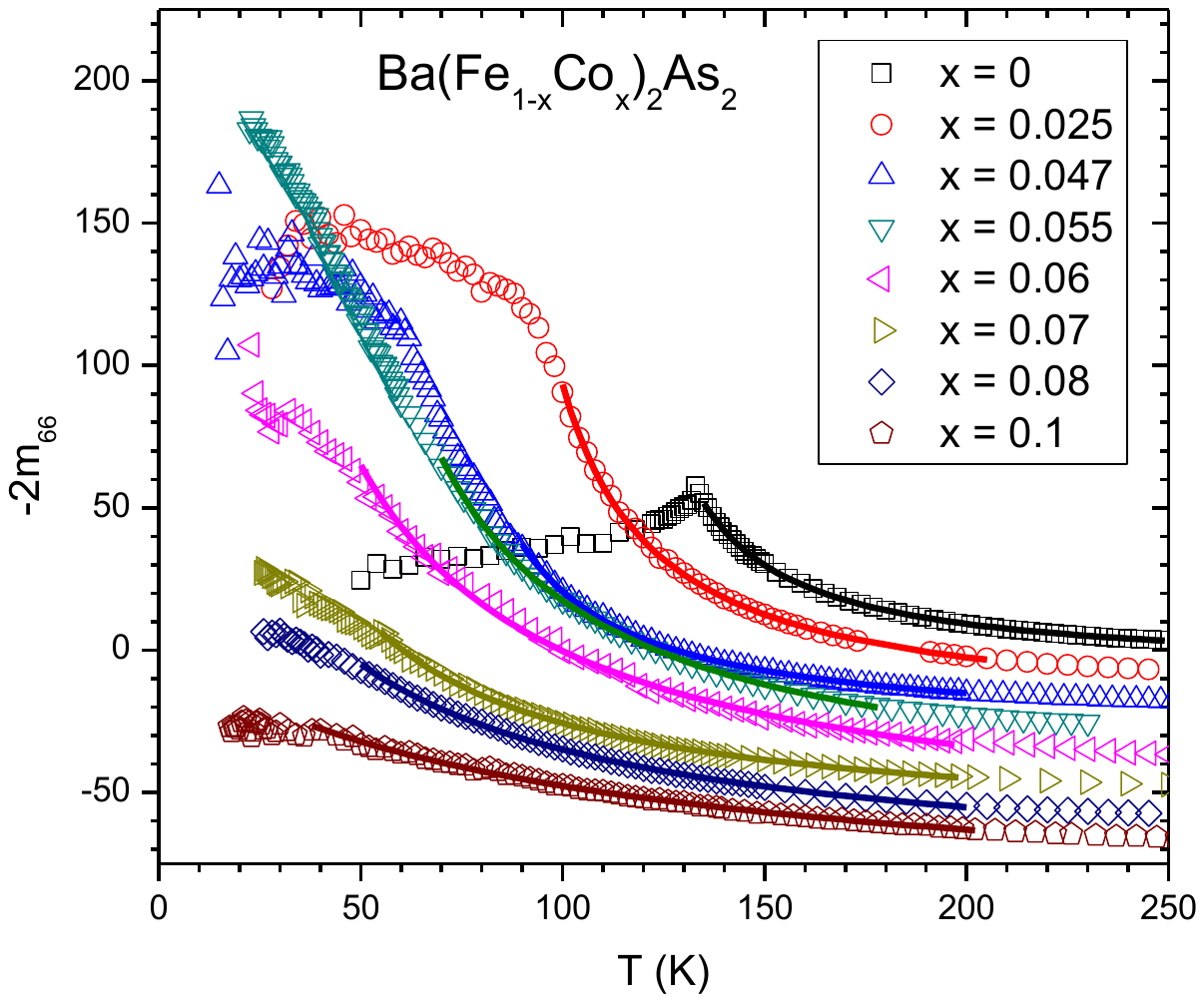}\label{FigS14}
\end{figure}
\noindent \textbf{Figure S14}: Temperature dependence of the $B_{2g}$ elastoresistance coefficient $2m_{66}$ of \CoBa122 for eight doping concentrations from $x = 0$ to $x = 0.1$. Solid curves shows fits to the Curie-Weiss behavior, and the fitting parameters are tabulated in table S1. Data were successively offset by 10 for clarity.

\clearpage

\begin{figure}\includegraphics[width=15cm]{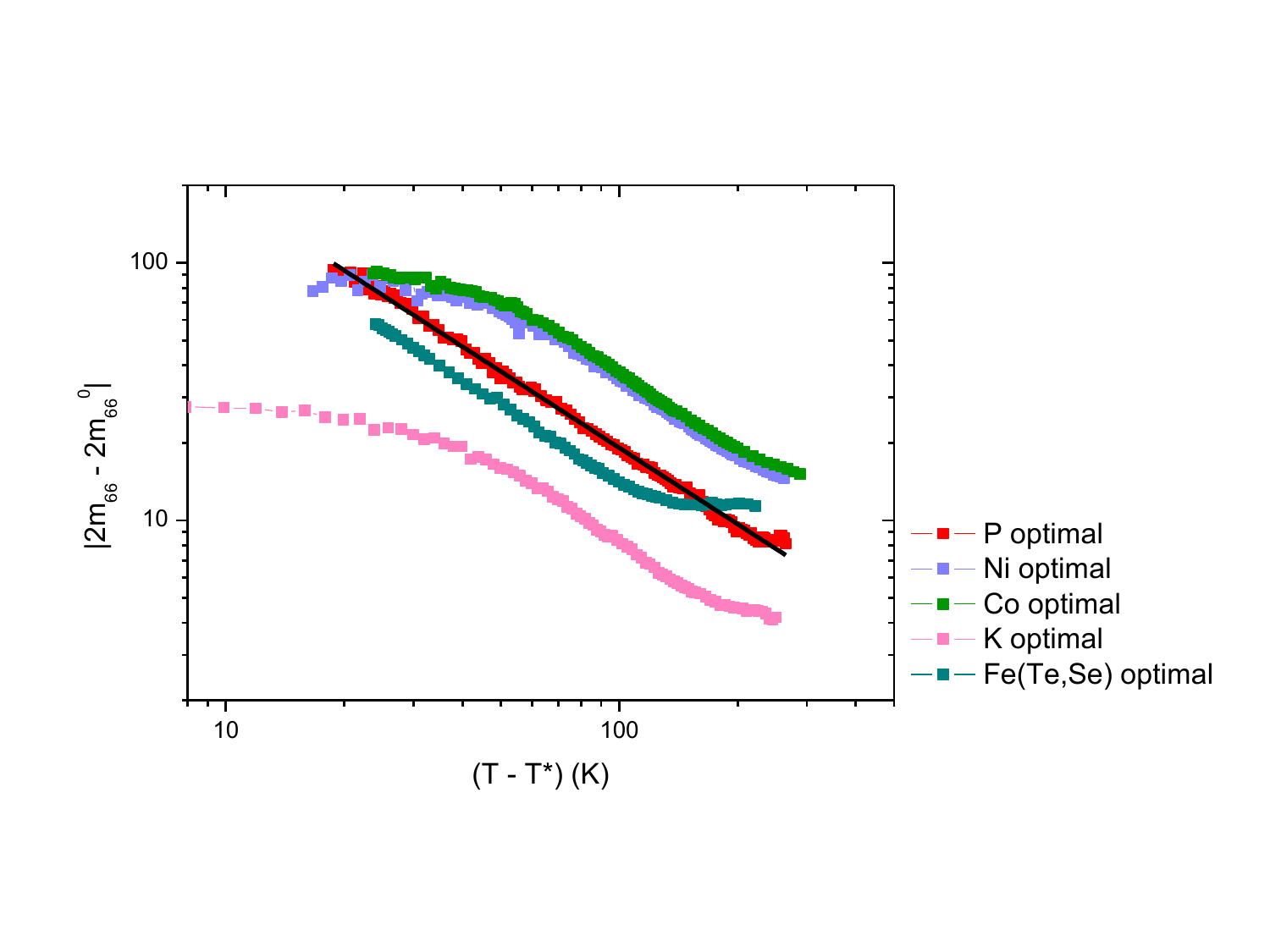}\label{FigS15}
\end{figure}
\noindent \textbf{Figure S15}: Power law behavior of the $B_{2g}$ elastoresistance of optimally doped BaFe$_2$(As$_{0.7}$P$_{0.3}$)$_2$. Figure shows $\log [|(2m_{66}-2m_{66}^0)|]$ v.s. $\log (T-T^*)$ for all measured optimally-doped iron-based superconductors. $2m_{66}^0$ and $T^*$ were extracted from the Curie-Weiss fit\cite{SOM}. Only the optimally P-substituted data (red) can be fitted linearly over the whole temperature range, with a slope $\gamma = 0.985 \pm 0.005$ (solid black line). Note that for Co, Ni, and P substitution, $2m_{66}-2m_{66}^0$ is negative, whereas for K substitution and FeTe$_{1-x}$Se$_{x}$, it is positive.

\clearpage
\begin{figure}\includegraphics[width=15cm]{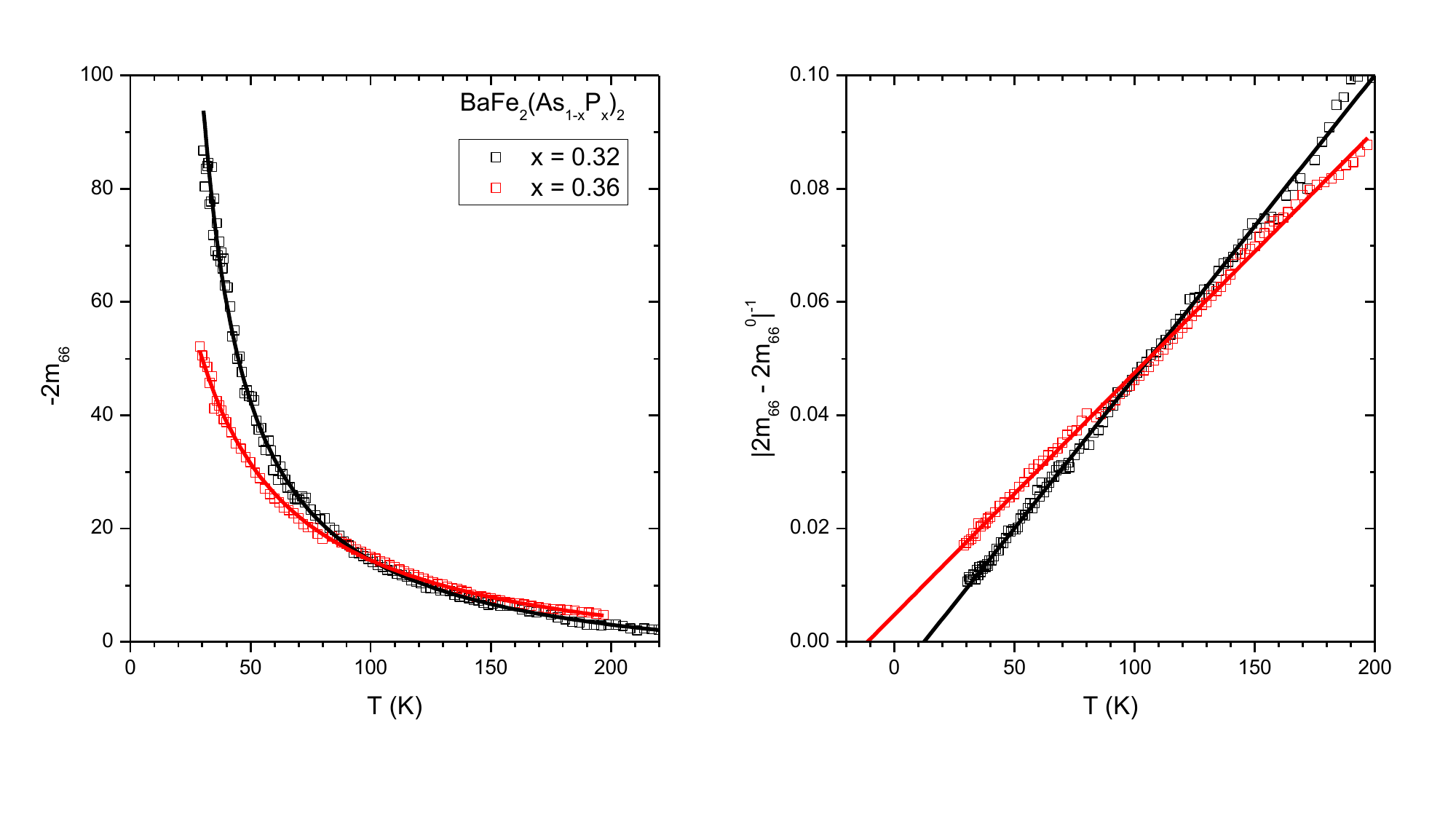}\label{FigS16}
\end{figure}
\noindent \textbf{Figure S16}: Temperature dependence of the $B_{2g}$ elastoresistance coefficient (a) $2m_{66}$ and its inverse (b) $|(2m_{66}-2m_{66}^0)|^{-1}$ of BaFe$_2$(As$_{1-x}$P$_{x}$)$_2$ for $x = 0.32$ and $x = 0.36$. Solid curves shows fits to the Curie-Weiss behavior, and the fitting parameters are tabulated in table S1. From, the x-intercepts of the linear fits of $|(2m_{66}-2m_{66}^0)|^{-1}$ it can be seen that the Weiss temperature crosses zero as the doping concentration increases from optimal doped to slightly overdoped. 
\end{widetext}


\begin{thebibliography}{10}
\bibitem{constant}There are several different ways to define a nematic order parameter in terms of equilibrium electronic correlation functions. Correspondingly, there is an ambiguity in defining the nematic susceptibility. Thus the constants $c$ and $c'$ not only depend on microscopic details, but they also depend on the precise microscopic definition we give to the nematic order parameter.
\bibitem{Strain} Typical strains using this method are of order $10^{-3}\sim10^{-4}$.
\bibitem{Dimension}  While most of the samples used in this paper are in the size range of full strain transmission, it is important to note that the phosphorus substituted iron arsenide samples are $\sim300\mu$m square and are in the regime of smaller strain transmission.
\end{thebibliography}
\end{document}